\documentclass[preprint2]{proto}
\usepackage{natbib}
\usepackage{times}
\usepackage{graphicx}
\usepackage{epstopdf}

\def\lsun{\hbox{L$_\odot$}}

\def\msun{\hbox{M$_\odot$}}

\def\t4{\hbox{t$_{\rm 4}$}}

\def\hii{\hbox{H$_{\rm II}$}}

\newcommand{\pc}{\mbox{{\rm pc}}}
\newcommand{\kms}{\mbox{{\rm km~s}$^{-1}$}}

\begin{document}

\title{\textbf{\LARGE The Formation and Early Evolution of Young Massive Clusters}}

\author {\textbf{\large Steven N. Longmore}}
\affil{\small\em Liverpool John Moores University}

\author {\textbf{\large J. M. Diederik Kruijssen}}
\affil{\small\em Max-Planck Institut f\"{u}r Astrophysik}

\author {\textbf{\large Nate Bastian}}
\affil{\small\em Liverpool John Moores University}

\author {\textbf{\large John Bally}}
\affil{\small\em University of Boulder, Colorado}

\author {\textbf{\large Jill Rathborne}}
\affil{\small\em CSIRO Astronomy and Space Science}

\author {\textbf{\large Leonardo Testi}}
\affil{\small\em European Southern Observatory}

\author {\textbf{\large Andrea Stolte}}
\affil{\small\em Argelander Institut f\"ur Astronomie, Bonn}

\author {\textbf{\large James Dale}}
\affil{\small\em Excellence Cluster Universe}

\author {\textbf{\large Eli Bressert}}
\affil{\small\em CSIRO Astronomy and Space Science}

\author {\textbf{\large Joao Alves}}
\affil{\small\em University of Vienna}

\begin{abstract}
\baselineskip = 11pt
\leftskip = 0.65in 
\rightskip = 0.65in
\parindent=1pc {\small We review the formation and early evolution of the most massive ($>$ few 10$^4\,\msun$) and dense (radius of a few pc) young stellar clusters, focusing on the role that studies of these objects in our Galaxy can play in our understanding of star and planet formation as a whole. Comparing the demographics of young massive cluster (YMC) progenitor clouds and YMCs across the Galaxy shows that gas in the Galactic Center can accumulate to a high enough density that molecular clouds already satisfy the criteria used to define YMCs, without forming stars. In this case formation can proceed ``in situ" -- i.e. the stars form at protostellar densities close to the final stellar density. Conversely, in the disk, the gas either begins forming stars while it is being accumulated to high density, in a ``conveyor belt" mode, or the time scale to accumulate the gas to such high densities must be much shorter than the star formation timescale. The distinction between the formation regimes in the two environments is consistent with the predictions of environmentally-dependent density thresholds for star formation. This implies that stars in YMCs of similar total mass and radius can have formed at widely different initial protostellar densities. The fact that no strong, systematic variations in fundamental properties (such as the IMF) are observed between YMCs in the disk and Galactic Center suggests that, statistically speaking, stellar mass assembly is not affected by the initial protostellar density. We then review recent theoretical advances and summarize the debate on three key open questions: the initial (proto)stellar distribution, infant (im)mortality and age spreads within YMCs. We  conclude that: (i) the initial protostellar distribution is likely hierarchical; (ii) YMCs likely experienced a formation history that was dominated by gas exhaustion rather than gas expulsion; (iii) YMCs are dynamically stable from a young age; and (iv) YMCs have age spreads much smaller than their mean age.  Finally, we show that it is plausible that metal-rich globular clusters may have formed in a similar way to YMCs in nearby galaxies. In summary, the study of YMC formation bridges star/planet formation in the solar neighborhood to the oldest structures in the local Universe.  \\~\\~\\~}

\end{abstract}

\section{\textbf{Overview of YMCs and their role in a global understanding of star and planet formation}}

\subsection{Motivation}

As exemplified by the reviews in this volume, the basic theoretical framework describing the formation of isolated, low-mass stars is now well established. This framework is underpinned by detailed observational studies of the closest star-forming regions. But how typical is the star and planet formation in Taurus, Perseus, or even Orion compared to the formation environment of most stars across cosmological timescales? 

The fact that half of the star formation in the Galaxy is currently taking place in the 24 most massive giant molecular clouds \citep{lee12}, suggests that even in the Milky Way at the present day, star formation regions in the local neighborhood are not typical. Environmental conditions were likely even more different at earlier epochs of the Universe. The epoch of peak star formation rate density is thought to lie between redshifts of 2 and 3 \citep{madau98,hopkins06b,bouwens11,moster13}, when the average gas surface density (and hence inferred protostellar densities) were significantly higher \citep[see e.g.][]{carilliwalter13}. Given that most stars in the solar neighborhood are at a similar age to the Sun \citep[$\sim$5\,Gyr: ][]{nordstrom04}, does it then make sense to compare the planetary populations observed around these stars to the protoplanetary disks in nearby star forming regions? 

The fundamental question underlying this line of reasoning is, ``Does the process of stellar and planetary mass assembly care about the environment in which the stars form?" If the answer is ``No", then studying the nearest star formation regions will tell us all there is to know about star and planet formation. If the answer is ``Yes", it is crucial to understand how and why the environment matters. The potential implications of these answers provide a strong motivation for comparing the process of star and planet formation in extreme environments, with that in nearby, well-studied, less extreme star formation regions.

\subsection{Young Massive Clusters: ideal probes of star and planet formation in extreme environments}
\label{sub:ymc_ideal_probes}

Stars are observed to form in a continuous range of environments and densities, from isolated molecular gas clouds expected to form single low-mass stellar systems \citep[e.g. B68:][]{alvesladalada2001}, through to giant molecular cloud complexes expected to form hundreds of thousands of stars across the full stellar mass range (see Molinari et al. this volume). Similarly, young stellar systems are found at a continuous range of mass and stellar densities \citep{bressert10}. Given this apparently continuous distribution in mass and density of both gas and stars, what motivates a definition for a distinct class of stellar systems beyond mere phenomenology?

The answer lies in the fact that most stellar systems dissolve shortly after forming, thereby feeding the field star populations of galaxies \citep{lada03}. Only a small fraction are simultaneously both massive and dense enough to remain gravitationally bound long after their formation and subsequent removal of the remaining natal molecular gas cloud. Being able to study an ensemble single stellar population long after formation offers many advantages, not least of which is the ability to retrospectively investigate the conditions under which the stars may have formed.

The most extreme examples of such stellar systems are globular clusters, which formed at the earliest epochs of the Universe and survive to the present day \citep{brodie06}. One of the ground-breaking discoveries of the Hubble Space Telescope was that massive stellar clusters, with properties that rival those found in globular clusters in terms of mass and stellar density, are still forming in the Universe today \citep{holtzman92}. These young massive clusters (YMCs) have stellar masses and densities orders of magnitude larger than typical open clusters and comparable to those in globular clusters. Crucially, they are also gravitationally bound and likely to be long-lived. As such, these stellar systems are potentially local-universe-analogs of the progenitors of globular clusters. 

At the same time, the apparent continuum of young cluster properties \citep[e.g.][]{bressert10}, suggests that YMCs merely represent extreme examples of their less massive and dense counterparts -- open clusters. As such, characterizing and understanding how YMCs form is critical to help make the connection between the range of physical conditions for star and planet formation between Galactic and extra-galactic cluster formation environments. 

\subsubsection{YMCs: definition and general properties}

In their recent review,  \citet{portegieszwart10} defined YMCs as stellar systems with mass $\gtrsim10^4$\,M$_\odot$ and with ages less than 100\,Myr but substantially exceeding the current dynamical time (the orbital time of a typical star). While the ultimate longevity of a stellar system will depend on the environment it experiences over time \citep{spitzer58}, this last criterion effectively distinguishes between presently bound clusters and unbound associations (see $\S$~\ref{sub sub:dist_young_stars} for further details). 

Given previous confusion in the literature caused by loose and varied definitions of what constitutes a stellar cluster (see $\S$~\ref{sub sub:dist_young_stars}), it is important to point out the implications of the above criteria. Firstly, there are many, well-known, massive associations of stars which do not pass these criteria (e.g the Cygnus OB association and the Orion Nebular Cluster). Secondly, YMCs that do pass these criteria (e.g. Westerlund 1, NGC 3603, Trumpler 14) may lie within a much larger stellar association, which as a whole does {\bf not} pass these criteria.  This is a direct consequence of the `continuum' of stellar properties discussed above. We emphasize that the focus here is on the YMCs and not the  more distributed stellar populations. This will impact the discussion in $\S$3 on gas expulsion, longevity and the presence of age spreads within clusters.

To date, nearly one hundred YMCs have been discovered in the local group and out to distances of a few Mpc. The properties of many of these are catalogued in \citet{portegieszwart10}. For convenience, we summarize their characteristic properties below. YMCs typically have radii of $\sim$1pc and core stellar densities $\geq$10$^3$\,M$_\odot$pc$^{-3}$. They are generally spherical, centrally-concentrated and often mass segregated (i.e. more massive stars are preferentially found towards the center of the cluster). The initial cluster mass distribution is not trivial to measure, but over many orders of magnitude in mass appears to be reasonably well approximated by a power law, $dN/dM \propto M^{-2}$, across all environments.  YMCs are found predominantly in starburst galaxies and mergers -- a couple of thousand are known to exist in the Antennae and NGC 3256, for example. These YMCs are typically more massive than those found in the Local Group and Milky Way. In the local universe (i.e. not starbursts/mergers), YMCs are typically found in the disks of galaxies. Globular clusters are predominantly found in galactic halos. Rotation has been observed in one YMC \citep[R136:][]{henault12b} as well as two intermediate age massive, dense clusters \citep[GLIMPSE-C01, NGC 1846:][]{davies11,mackey13}. Given the difficulty in measuring rotation, it is currently unknown how common this property is among YMCs.

\subsubsection{The role of YMCs in the broader context of planet, star and cluster formation}

The properties of YMCs make them ideal probes of star and planet formation in extreme environments. Stars forming at such high (proto)stellar densities suffer the maximal effects of feedback from surrounding stars. Also, the very short dynamical time increases the likelihood of interactions with nearby stars at all stages of the formation process. Therefore, studying the formation of stars within a YMC compared to low stellar density systems, offers an opportunity to quantify how dynamical encounters and stellar feedback affect the process of stellar mass assembly.

YMCs contain a very large number of stars of a similar age (age spreads $\lesssim$1\,Myr: see $\S$~\ref{subsub:age_spreads}). These stars likely formed from the same gas cloud, so were born in the same global environmental conditions and have the same chemical composition. This makes YMC precursor gas clouds the perfect test beds to study the origin of the stellar initial mass function (IMF). For example, by studying YMC progenitor clouds before the onset of star formation, it should be possible to determine if the final stellar mass is set by the initial mass distribution of gas fragments, or alternatively, by these initial fragments subsequently accreting unbound gas from the surrounding environment (see the review by Offner et al., this volume for a more detailed discussion on the origin of the IMF).

YMC precursor clouds are also, statistically-speaking, the best place to search for the progenitors of the most massive stars. While the progenitors of many late-O and early-B stars have been identified, precursors to stars destined to be hundreds of solar masses still prove elusive. Identifying such precursors will help in the theoretical challenge to understand how the most massive stars form (see Tan et al. this volume for a review on high-mass star formation).

YMCs occupy a unique position in understanding cluster formation. As a bridge in the apparent continuum of cluster mass and density distributions between open and globular clusters, studying their global and environmental properties can provide insight into what conditions are required in order for bound clusters to form. Is there a single, scalable formation mechanism applicable to all clusters? Or are additional mechanisms required to accumulate such a large gas mass in a small volume for the most massive clusters? YMCs may be used as a direct probe to understand the conditions required for globular cluster formation.

\subsection{Scope of the review}

Several fundamental, unanswered questions about the formation and early evolution of YMCs currently limit their utility as probes of star and planet formation in extreme environments. For example, while the spatial distribution of stars in YMCs older than a few Myr is relatively well known \citep{king66,elson87}, it is not clear how this relates to the initial protostellar or gas distribution \citep[e.g.][]{rolffs11}. Any initial substructure that existed in the gas and protostars is erased quickly \citep{mcmillan07}. Therefore, if the stars actually formed at a much lower density -- and hence in a much less extreme environment than assumed from the present-day stellar density -- and then grew into a massive, dense cluster over time, there would be little evidence of this in the final stellar surface density distribution as the structure would have been erased by violent relaxation. A potential, new method of deriving the initial conditions of cluster formation a posteriori would be to consider quantities that are conserved during violent relaxation, such as the degree of mass segregation, and to combine these with a measure of the remaining substructure. Collapsing, virialised and unbound stellar structures may follow distinct evolutionary histories in the plane defined by these quantities \citep{parker13b}.

More generally, it is not clear if all clusters of the same mass and radius form from gas with similar properties. Are there different ways to form bound clusters of similar final stellar properties? If so, and if stellar mass assembly depends on the protostellar environment, it is important to understand how and when these different mechanisms operate.

Understanding these questions requires making the link between the evolution of the initial progenitor gas clouds towards the final, gas-free stellar populations. However, the properties of YMCs have been derived almost exclusively from optical/infrared observations. This has strongly biased YMC detection towards clusters with relatively low extinctions (A$_{\rm v}\lesssim 30$), preferentially selecting clusters which are already gas-free -- i.e. older than a (few) Myr. This bias means that very little is known about YMCs younger than this, or their progenitor gas clouds. 

In this review we focus on: (i) the initial conditions of proto-YMCs, (ii) the gas-rich, first (few) $\sim$Myr in the life of YMCs as they are forming stars, and (iii) the evolution shortly thereafter. This is intended to complement the \citet{portegieszwart10} review, which focussed on the aspects of YMCs older than a few Myr.

\section{\textbf{Molecular cloud progenitors of YMCs -- the Initial Conditions}}

Understanding the formation of YMCs requires first finding samples of YMC progenitor clouds that represent the initial conditions (i.e. before star formation commences), which can be directly compared to their more evolved stellar counterparts. However, very few pre-star-forming YMC progenitor clouds have been identified. In an attempt to understand the plausible range of properties for the initial molecular cloud progenitors of YMCs, we consider some simplified formation scenarios below. 

\subsection{Simplified YMC formation scenarios}
\label{sub:simple_formation_scenarios}

The most basic initial condition for YMC formation is a gas reservoir with enough mass, M$^{\rm init}_{\rm gas}$, to form a stellar cluster of mass, M$_*\geq 10^4$\,M$_\odot$. These two quantities are trivially related via the star formation efficiency, $\epsilon$, through M$^{\rm init}_{\rm gas}=$ M$_*$/$\epsilon$. To span the expected range of molecular cloud progenitor properties, we investigate two extremes in the initial spatial distribution of the gas (i.e. before the onset of any star formation) relative to that of the final stellar cluster. Firstly, we consider the case where the initial extent of the gas, R$^{\rm init}_{\rm gas}$, equals that of the resulting cluster, $R_*$. Then we consider the case where R$^{\rm init}_{\rm gas}$ is substantially (factors of several or more) larger than $R_*$.

\subsubsection{R$^{\rm init}_{\rm gas}=R_*$ : ``in-situ formation"}

In this scenario, all the required gas is gathered into the final star cluster volume \emph{before} star formation commences (i.e. in-situ star formation). In principle, a direct observational prediction of this model would be that one would expect to find gas clouds with mass M$_{\rm gas}$ and radius R$_*$ with no signs of active star formation. However, the probability of finding such a cloud under this scenario depends on the ratio of the time taken to accumulate the gas within the final cluster volume to the time taken for star formation to get underway there. The very high densities required to form a YMC implies that the gas inside the final cluster volume will have a correspondingly short free-fall time in this scenario. If star formation happens on a dynamical timescale, this implies that either the time taken to accumulate the gas reservoir there must also be very short, or that star formation inside the final cluster volume is somehow delayed or suppressed while the gas accumulates.

In the former case, which we term `in-situ fast formation', the accumulation time is very short and star formation is almost instantaneous once the gas is accumulated. It is therefore very unlikely that a YMC progenitor cloud with mass M$_{\rm gas}$ and radius R* with no signs of active star formation would be observed, but significant numbers of such clouds exhibiting ongoing star formation should be observed.
In the latter case, dubbed `in-situ slow formation', the accumulation time is long and star formation is delayed until most of the mass required to build the YMC has accumulated inside R*. Significant numbers of clouds with mass close to M$_{\rm gas}$ and radius R* but with no active star formation would therefore be observed in this case.

\subsubsection{R$^{\rm init}_{\rm gas} > R_*$ : ``conveyor belt formation"}

In this scenario, the gas that eventually ends up in the YMC is initially much more (factors of several or greater) extended than that of the final cluster. The initial, quiescent gas clouds begin forming stars at a much lower global surface/volume density than in the previous ``in-situ" scenario. In order for the proto-cluster to reach the required final stellar densities, the gas and forming stars must converge into a bound stellar system. The most likely agents to enhance gas density are the convergence of two initially independent gas flows, or the gravitational collapse of a single cloud. In this scenario, one would never expect to see clouds of M$^{\rm init}_{\rm gas}$ and R$_*$ with no signs of active star formation. 

As outlined in section $\S$\ref{sub:theory}, the long-term survival of the cluster is strongly influenced by the mechanism and timescale for gas removal. The time for gas dispersal, $t_{\rm disp}$, therefore places a strong upper limit on the time during which it is possible to form a cluster through convergence/collapse. Given a convergence velocity, $V_{\rm conv}$, this sets an upper limit to the initial radius of the gas to be R$^{\rm init}_{\rm gas} =  $ R$_* +$ V$_{\rm conv} t_{\rm disp}$.

The timescale for star formation and the observed age spreads are key diagnostics for distinguishing between these scenarios. We look at the observational evidence for variation in these properties in YMCs in $\S$\ref{sub:theory}.

\subsection{Comparing YMC and progenitor cloud demographics}
\label{sub:comp_demographics}

We now demonstrate how one can use the observed demographics of molecular cloud populations, compared with those of the stellar cluster populations in the same region, to test these different YMC formation scenarios. 

Firstly, we assume that in a region with a large enough volume to sample all stages of the star/cluster formation process, the present day molecular cloud population will create similar clusters as those observed at the present day. In practice this implies that the star formation rate, cluster formation rate and the distribution of stars into clusters of a given mass and density should have been constant over several star formation life cycles. This seems a reasonable assumption for disks in nearby galaxies, but may not hold in mergers, starburst systems or dwarf galaxies \citep[see][]{kruijssen13e}.

The most massive gas clouds (M$^{\rm max}_{\rm gas}$) seem the obvious birth sites for the most massive clusters (M$^{\rm max}_*$). If no existing gas clouds have enough mass to form the observed most massive clusters (i.e. M$^{\rm max}_{\rm gas}\ll$ M$^{\rm max}_*/\epsilon$), these clouds must gain additional mass from elsewhere (e.g. through merging gas flows or accreting lower density gas from outside the present-day boundary) -- i.e. ``conveyor belt" formation.

On the other hand, if there are gas clouds of sufficient mass (i.e. M$^{\rm max}_{\rm gas}\geq$ M$^{\rm max}_*/\epsilon$), then the spatial/kinematic sub-structure of this gas and the distribution of star formation activity within these clouds can provide a key to the formation mechanism. If concentrations of gas with M$_{*}/\epsilon$ within $\sim$R$_{*}$ exist, then finding a sizable fraction with no star formation activity would indicate YMCs are forming ``in-situ". If the gas in the most massive clouds is spatially distributed such that no sub-region of any cloud contains a mass concentration of M$_{*}/\epsilon$ within $\sim$R$_{*}$, then in-situ formation seems highly unlikely. In which case the stars forming in the gas must converge and become gravitationally-bound before the star formation can disrupt the cloud. Evidence for such convergence should be imprinted in the gas kinematics, e.g.  velocity dispersions of order V$_{\rm conv} =$~(R$^{\rm init}_{\rm gas}-$~R$_*$)/$t_{\rm disp}$. Inverse P-Cygni profiles and red/blue-shifted line profile asymmetries may also be observed but care must be taken interpreting such spectral line diagnostics \citep[][]{smith12, smith13}.

\subsubsection{Observational tracers and diagnostics}
\label{subsub:obs_tracers_diagnostics}

We now investigate the feasibility of directly comparing YMC and progenitor cloud demographics given current observational facilities. A fundamental limitation is the distance to which it is possible to detect a precursor cloud of a given mass. ALMA's factor $>$10 improvement in sensitivity compared to existing (sub)mm interferometers makes it the optimal facility for detecting gas clouds out to large distances. At a frequency of 230\,GHz (wavelength of 1.26\,mm) ALMA will achieve a 10$\sigma$ continuum sensitivity limit for a one hour integration of approximately 0.1\,mJy  (assuming 8\,GHz bandwidth). Assuming gas and dust properties similar to those in massive star forming regions in the Milky Way (gas temperature of 20K, gas:dust ratio of 100:1,  \citet{ossenkopf_henning94} dust opacities) this sensitivity limit corresponds very roughly to a mass limit of \{10$^5$, 10$^7$\}\,M$_\odot$ at a distance of \{0.5, 5\}\,Mpc. This simplistic calculation neglects several subtleties (e.g. the effects of beam dilution, higher gas temperatures in vigorously star forming systems and metallicity variations on the gas-to-dust ratio and dust opacity). However, it illustrates that the gas cloud populations previously accessible within the LMC/SMC can now be probed out to M31/M33 distances, and similar studies currently being done on M31/M33 GMC populations will be possible out to more extremely star forming galaxies like M82 and NGC253.

Emission from the CO molecule is another standard tracer of GMC populations. A combination of the excitation conditions and abundance means for a gas cloud of a given mass, low J transitions of CO are usually brighter than the dust emission in extragalactic observations. This means that the mass estimates above provide a lower mass limit to the detectability of gas clouds in CO. 

However, the expected high volume and column densities of YMC progenitor clouds means that CO may not be the ideal molecular line tracer for identification purposes. To illustrate this point, we note that a fiducial YMC progenitor cloud of 10$^5$\,M$_\odot$ with radius 1\,pc (e.g. as would be expected to form a 3$\times$10$^4$\,M$_\odot$ cluster through in-situ formation, assuming a 30\% star formation efficiency) would have an average volume and column density of $2\times10^4\,\msun\,$pc$^{-3}$ ($4\times10^5$\,cm$^{-3}$) and $3\times10^4\,\msun\,$pc$^{-2}$ ($2\times10^{24}$\,cm$^{-2}$), respectively. This column density corresponds to a visual extinction of $\sim$2000\,mag. At such high densities, even if observations can resolve the gas emission down to parsec scales, the CO emission will be optically-thick. Therefore such observations can only probe a $\tau=1$ surface, not the bulk of the gas mass. Similar resolution observations of molecular transitions with a higher critical density (comparable to that of the average volume density in the YMC progenitor cloud) are required to pinpoint these clouds. As an interesting aside,  such high column densities render H$\alpha$ -- the traditional SF indicator in extragalactic observations -- completely unusable. Probing gas clouds with and without prodigious embedded star formation activity will therefore rely on complementary observations to measure star formation tracers less affected by extinction (e.g. cm-continuum emission to get the free-free luminosity, or far-IR observations to derive the bolometric luminosity).

The gas mass inferred from observations is a beam-averaged quantity. In other words, if a gas cloud is much smaller than the observational resolution and sits within a lower density environment, the measured beam-averaged column/volume density will be lower than the true value, leading to incomplete YMC progenitor samples. However, even when not operating at its best resolution, ALMA should routinely resolve the $\sim$\,pc-scale YMC progenitor cloud sizes out to several tens of Mpc. 

To measure what influence the high protostellar density environment has on forming protostars and their planetary systems, it is necessary to resolve individual stellar systems. In practice the projected protostellar separation will vary, both from source to source, and as a function of radius within an individual region. However, relying on the fact that the average core mass is $\sim$1\,$\msun$, the characteristic projected separation of protostars within a protocluster of mass M$_*$ and radius R$_*$ is proportional to R$_*$\,(M$_*/\msun)^{-1/2}$. The typical projected angular separation of protostars within a protocluster as a function of distance to the protocluster, D, is (very roughly) 4\,(R$_*$/pc)\,(M$_*$/$10^4$\msun)$^{-1/2}$\,(D/kpc)$^{-1}$\,arcsec. Even at the maximum resolution of ALMA of 0.01\arcsec (i.e. using the most extended 10\,km baselines at the highest frequency [Band 9]), it will not be possible to resolve individual stellar systems in YMC progenitor clouds beyond about 100\,kpc (i.e. LMC and SMC distances). The maximum angular resolution limit for ALMA is comparable to that expected from future 30$-$40\,m aperture optical/infrared telescopes. For \emph{at least} the next several decades, observations probing the physics shaping the IMF in dense stellar systems must be limited to star forming regions in the LMC/SMC and closer. 

Assuming it is possible to resolve individual protostellar systems, the observational limit then becomes one of mass sensitivity. Even with ALMA and choosing the closest possible targets in the Galaxy, deep integrations will be required to probe the gas expected to form stars across the full stellar mass range.

Understanding the gas kinematics across a range of densities and spatial scales is necessary to distinguish between the different formation scenarios of YMCs.  The `conveyor belt' model, for example, suggests that large amounts of low- or moderate-density gas should be rapidly infalling. Given the new frontier in sensitivity being opened up by ALMA, it is not clear at this stage what the best transitions for this purpose might be. However, studies of (less extreme) massive and dense high-mass star forming regions are paving the way \citep[e.g.][]{peretto13}. Deriving the spatial and kinematic distribution of mass as a function of size scale will likely require simultaneously observing many different transitions to solve for opacity, excitation and chemistry variations. Extreme environments, like the Galactic Center, will prove especially challenging in this regard.

\subsection{YMCs and progenitor clouds in the Milky Way}

Extragalactic observations will be crucial to probe the formation of the most massive YMCs in a wide range of environments (e.g. galaxy mergers). However, the discussion in $\S$~\ref{subsub:obs_tracers_diagnostics} shows that for the foreseeable future the Milky Way, and to a lesser extent the LMC and SMC, are the only places in the Universe where it will be possible to resolve sites of individual forming protostellar/planetary systems in regions which have protostellar densities $>10^4\,\msun$\,pc$^{-3}$. This means they are also the only places where it will be possible to directly test the effect of extreme environments on individual protostellar systems. This provides a strong motivation to identify a complete sample of YMCs and their progenitor clouds in the Galaxy. Such a catalog does not yet exist due to the difficulty in finding clouds at certain stages of the formation process.

On the one hand, it is straightforward to find all the clouds in the Galaxy with embedded stellar populations $>10^4$\,M$_\odot$. Their high bolometric luminosity ($>10^6\,\lsun$) and ionizing flux (Q$>10^{51}$/s) make them very bright at far-IR wavelengths (where the spectral energy distribution peaks) and cm wavelengths (which traces the free-free emission from the ionized gas at wavelengths where the clouds and the rest of the Galaxy are optically-thin). As a result, these sources with prodigious embedded star formation have been known since the early Galactic plane surveys at these wavelengths \citep[e.g.][]{westerhout58}, and many such objects have been studied in detail \citep[e.g.][]{plume97, shridharan02, beuther02, lumsden13}.

The difficulty in generating a complete catalog of YMC progenitor clouds has been finding those before star formation has begun. At this early stage there is no ionizing radiation and the luminosity is low. Therefore, these regions do not stand out in cm or far-IR wavelength surveys. However, as discussed above, in all three scenarios they must have a large gas mass in a small volume. As such, they should be easy to pick out as bright, compact objects at mm and sub-mm wavelengths. However, observational limitations have meant that Galactic plane surveys at these wavelengths have only been possible over the last few years. Previous targeted surveys for young massive proto-clusters have not found any starless gas clouds with $>10^5$\,$\msun$ at parsec size-scales \citep[e.g.][]{faundez04, hill05, simon06, rathborne06, purcell06, peretto_fuller09}. 

However, thanks to a concerted effort from the observational Galactic star formation community over the last few years (see the review by Molinari et al., this volume), the data will soon be available to compile a complete list of YMC progenitor clouds in the Milky Way needed to make definitive statements about the relative populations of YMC progenitor clouds with and without prodigious star formation activity. To date, systematic, blind, large-area searches for YMC progenitor clouds at all stages of the cluster formation process have been published for two regions of the Galaxy: the first quadrant \citep{ginsburg12} and the inner 200\,pc \citep{longmore13b}.  In the near future, results from continuum surveys like ATLASGAL \citep{schuller09,contreras13} combined with spectral line studies \citep[e.g. MALT90, CHAMP, ThrUMMS, Mopra Southern Galactic Place CO survey -- ][]{foster11, jackson13, barnes11, burton13}, will extend the search to the fourth quadrant. For example, \citet{urquhart13} have already identified a sample of YMC candidates with signs of active star formation and Contreras et al. (in prep) have identified YMC candidates at all evolutionary stages through the MALT90 survey. In the longer-term, HiGAL \citep{molinari10} will provide a sensitive, uniform survey across the whole Galaxy. However, our analysis relies on having complete samples at all stages of the cluster formation process, so we focus on the extant surveys of the first quadrant and inner 200\,pc of the Galaxy below.

\subsubsection{The first quadrant of the Galaxy}
\label{subsub:first_quad}

\citet{ginsburg12} used BGPS data \citep{aguirre11} to carry out a systematic search for YMC progenitor clouds in the first Galactic quadrant, $l=6^\circ-90^\circ$ $|b|<0.5^{\circ}$.  This region is equivalent to $\sim$30\% of the total Galactic surface area, assuming an outer Galactic radius of 15\,kpc.  In this region \citet{ginsburg12} identified 18 clouds with mass M$_{\rm gas}>10^4\,\msun$ and radius r\,$\leq$\,2.5\,pc. All of these clouds have  gravitational escape speeds comparable to or larger than the sound-speed in photo-ionized gas, so pass the \citet{bressert12c} criteria for YMC progenitor clouds. Crucially, all 18 of these clouds are prodigiously forming stars. None of them are starless. \citet{ginsburg12} use this to place an upper limit of 0.5\,Myr to the starless phase for the clouds in their sample. This is similar to the upper limit on the lifetimes of clouds forming high-mass stars by \citet{tackenberg12}. Assuming a star formation efficiency of 30\%, only 3 of the 18 identified clouds are massive enough to form bound stellar clusters of $10^4$\,$\msun$.

\subsubsection{The inner 200\,pc of the Galaxy}
\label{subsub:inner_200pc}

\citet{longmore13b} conducted a systematic search for likely YMC progenitor clouds in the inner 200\,pc of the Galaxy by combining dust continuum maps with spectral line maps tracing molecular gas at high volume density. Based on maps of the projected enclosed mass as a function of radius, they identified 6 clouds as potential YMC progenitors. Intriguingly, despite having extremely high column densities (up to $\sim$10$^{24}$\,cm$^{-2}$; 2$\times10^4\,\msun$\,pc$^{-2}$) and being opaque up to 70\,$\mu$m, four of the six potential YMC progenitor candidates show almost no signs of star formation. The upper limit to the free-free emission from sensitive cm continuum observations, shows that there are, at most, a small number of early B stars in these four clouds \citep{immer12,rodriguez_zapata13}. This is in stark contrast to the clouds of similar mass and density seen in the disk of the Galaxy, which are all prodigiously forming stars (see $\S$~\ref{subsub:first_quad}).  

\subsubsection{Comparison of the 1$^{st}$ quadrant and inner 200\,pc}

Following the arguments outlined in $\S$~\ref{sub:comp_demographics}, if the molecular cloud population in a given region can be expected to produce the stellar populations in the same region, the cloud and stellar demographics can be used to infer something about the underlying formation mechanism. We now attempt this for the first quadrant and inner 200\,pc of the Galaxy. 

The first step is testing whether the assumption of the observed gas clouds producing the observed stellar populations holds for these regions. The region observed by \citet{ginsburg12} covers 30\% of the surface area of the Galaxy (assuming a Galactic radius of 15\,kpc). The inner few hundred pc of the Galaxy contains roughly 10\% of the molecular gas in the Galaxy \citep[see][for mass estimates]{pierce-price2000, ferriere07, kalberla_kerp09, molinari11}. If the star formation rate in these regions has remained constant over several star formation cycles, it seems reasonable to assume such large gas reservoirs will produce statistically similar stellar populations as observed at the present day. However, once a stellar system has formed, the environment in the Galactic Center is potentially a lot more disruptive than in the disk. Indeed, even dense clusters like YMCs are not expected to live longer than (or be detectable after) a few Myr in the Galactic Center \citep[e.g.][]{portegieszwart01,pz02b,kim_morris03}. For that reason, it is crucial to only compare the demographics of stellar clusters in the first quadrant and Galactic Center that are younger than this age.

We conclude that a reliable metric to investigate the different formation mechanisms is to compare the number of  YMCs younger than a few Myr (N$_{\rm YMC}$), to the number of YMC progenitor clouds with prodigious star formation activity (N$_{\rm cloud}^{\rm active}$), to the number of YMC progenitor clouds with no discernible star formation activity (N$_{\rm cloud}^{\rm no\, SF}$). In other words, the ratio of N$_{\rm YMC}$~:~N$_{\rm cloud}^{\rm active}$~:~N$_{\rm cloud}^{\rm no\,\,SF}$ contains information about the relative lifetime of these three stages. 

The inner 200\,pc of the Galaxy contains two YMCs -- the Arches and Quintuplet clusters (we exclude the nuclear cluster as this most likely has a different formation mode: see \citet{genzel10b} for a review) -- and two SF active clouds, Sgr~B2 and Sgr~C. Combined with the four quiescent clouds from $\S$~\ref{subsub:inner_200pc}, the N$_{\rm YMC}$~:~N$_{\rm cloud}^{\rm active}$~:~N$_{\rm cloud}^{\rm no\,\,SF}$ ratio in the inner 200pc of the Galaxy is then 2:2:4. 

Turning to the first quadrant, there is presently one known YMC in W49 \citep[][]{alves_homeier03}. Given the observational difficulties in finding unembedded YMCs at large distances through the Galactic disk, others may well exist. Completeness is not an issue for the two earlier stages (see $\S$~\ref{subsub:obs_tracers_diagnostics}). Combined with the number of SF active and quiescent clouds from $\S$~\ref{subsub:first_quad}, the  N$_{\rm YMC}$~:~N$_{\rm cloud}^{\rm active}$~:~N$_{\rm cloud}^{\rm no\,\,SF}$ ratio is then 1:3:0. 

Comparing the N$_{\rm YMC}$~:~N$_{\rm cloud}^{\rm active}$~:~N$_{\rm cloud}^{\rm no\,\,SF}$ ratios between the inner 200\,pc and first quadrant shows both regions are producing a similar number of YMCs with ages less than a few Myr. However, there is a large disparity between the number of progenitor clouds with/without star formation in the two regions. N$_{\rm cloud}^{\rm no\,\,SF}=0$ for the first quadrant but N$_{\rm cloud}^{\rm no\,\,SF}=4$ for the inner 200\,pc. Comparing to the predictions of the scenarios in $\S$~\ref{sub:comp_demographics}, the Galactic Center appears to be forming YMCs in an ``in-situ, slow formation" mode, whereas the disk appears to be consistent with a ``conveyor belt" or ``fast in-situ" mode of formation. 

In summary, studying the currently-available data in the Galaxy suggests that YMCs in different regions accumulate their mass differently. The two regions studied contain a sizeable fraction of the gas in the Milky Way, so it seems reasonable to conclude that this is representative of YMC formation as a whole in the Galaxy. Of course, when similar data becomes available for the rest of the Milky Way -- in particular the fourth quadrant which contains a large fraction of the gas in the Galactic disk -- it is important to test this result.

However, these Galactic regions only represent a small fraction of all the environments in the Universe known to be forming YMCs. Clearly it would be foolhardy at this stage to draw any general claims about YMC formation from a dataset sampling such a small fraction of the total number of regions forming YMCs. Future observational studies comparing the gas and stellar demographics across the full range of environments are required to make any such general, empirically-based statements about YMC formation. In the upcoming ALMA, JWST and ELT era, the datasets needed to solve this problem should become available.

In the mean time, we can still make progress in a general understanding of the YMC formation process from what we learn in the Galaxy if we can understand two key aspects: (i) if/how the underlying physical mechanism for YMC formation in the Galaxy depends on the environment, and, (ii) how those environmental conditions compare to other YMC-forming environments across cosmological timescales.

\subsection{The role of the environment for YMC formation}

We now investigate how differences in the environmental conditions may be playing a role in YMC formation. Following from the previous discussion, we start by comparing the general properties of the gas in the Galactic Center and the disk, before focussing on the properties of individual YMC progenitor clouds in the two regions.

\subsubsection{Comparison of gas properties across the Milky Way}
\label{sub sub:MW_gas_properties}

The general properties of the gas in the disk and the center of the Galaxy are both well characterized, and are known to vary substantially from each other \citep[for reviews see Molinari et al this volume;][]{morris96,ferriere07}.  In summary, the gas in the Galactic Center lies at much higher column and volume density \citep{longmore13}, has a much larger velocity dispersion at a given physical size \citep{shetty12} and has a higher gas kinetic temperature \citep{ao13,mills13}. The offset between the gas and dust temperature \citep{molinari11} in the Galactic Center is thought to be either due to the orders of magnitude larger cosmic ray flux than in the disk, or the widespread shocks observed throughout the gas \citep{ao13,yusefzadeh13}. The disk has a well-known metallicity gradient with galactocentric radius of $-0.03$ to $-0.07$ dex kpc$^{-1}$ \citep{balser11}. The metallicity in the Galactic Center is measured to vary within a factor of two of the solar value \citep{shields_ferland94,najarro2009}.

There is evidence that a combination of the environmental factors and the global properties of the gas leads to differences in how the star formation proceeds between the two regimes. Given the large reservoir of dense gas in the Galactic Center, the present-day star formation rate is at least an order of magnitude lower than that predicted by star formation relations where the star formation scales with the gas density \citep[e.g.][]{lada12,krumholz12a, beuther12, longmore13, kauffmann13}. \citet{kruijssen13c} find that the currently low SFR in the Galactic Center is consistent with an elevated critical density for star formation due to the high turbulent pressure. They propose a self-consistent cycle of star formation in the Galactic Center, in which the plausible star formation inhibitors are combined. However, the fact that (i) there is a non-zero star formation rate in the Galactic Center (albeit an order of magnitude lower than predicted given the amount of dense gas), and (ii) at least two YMCs are found in the Galactic Center, means that some mechanism must be able to overcome any potential suppression in star formation in a small fraction of the gas. As the details of this mechanism are of potential interest in understanding why the YMC formation mode in the Galactic Center may be different from the disk, in $\S$~\ref{subsub:ymc_form_gc} we examine this further before turning in $\S$~\ref{subsub:ymc_form_disc} to YMC formation in the disk. 

\subsubsection{YMC formation in the Galactic Center}
\label{subsub:ymc_form_gc}

A global understanding of star formation in the Galactic Center is hampered by the difficulty in determining the 3D distribution of the gas and stars. Building on earlier efforts \citep[e.g.][]{binney91}, \citet{molinari11} put forward a model that the ``twisted ring" of dense molecular gas of projected radius $\sim$100\,pc that they identified as very bright sub-mm continuum emission in the HiGAL data, was on elliptical X2 orbits (i.e. orbits perpendicular to the long axis of the stellar bar). In this scenario, the two prominent sites of star formation in the ring -- Sgr~B2 and Sgr~C -- lie at the location where the X2 orbits intersect with the X1 orbits (i.e. gas streams funneled along the leading edge of the stellar bar from the disk to the Galactic Center). In this picture, the collision of dense gas clouds may lead to YMC formation \citep[see e.g.][]{stolte08}.

Based on the observed mass distribution and kinematics, \citet{longmore13b} postulated that the gas in this ring may be affected by the varying gravitational potential it experiences. They hypothesized that the net effect of the interaction is a compression of the gas, aided by the gas dissipating the tidally-injected energy through shocks. If the gas was previously sitting close to gravitational stability, the additional net compression of the gas might be enough for it to begin collapsing to form stars. If this hypothesis proves correct, one can use the known time since pericentre passage to effectively follow the physics shaping the formation of the most massive stellar clusters in the Galaxy, and by inference the next generation of the most massive stars in the Galaxy, as a function of {\bf absolute} time. Numerical simulations by several different groups show that this scenario is plausible and can accurately reproduce the observed gas properties (Lucas \& Bonnell in prep, Kruijssen, Dale, Longmore et al in prep). The fact that the gas in this region has already assembled itself into clouds of $\sim10^5\msun$ and radius of a few pc before any star formation has begun, suggests that once the gas becomes gravitationally bound, it will form a young massive cluster.

The extreme infrared-dark cloud, G0.253$+$0.016 \citep[M0.25, the ``Lima Bean", the ``BrickÓ: ][]{lis94, lismenten98, bally10, longmore12} is the best studied example of such a cloud. Despite containing $\sim$10$^5\,\msun$ of gas in a radius of $\sim$3\,pc, the only signs of potential star formation activity are one 22\,GHz H$_2$O maser \citep{lis94} and several compact radio sources at its periphery \citep{rodriguez_zapata13}, indicating that at most a few early B-stars have formed. As expected, the average gas column density is very high \citep[$>10^{23}$\,cm$^{-3}$;][]{molinari11}. \citet{kauffmann13} showed that at 0.1pc scales, there are very few sub-regions with high column density contrast (corresponding to densities 2$\times$10$^5$\,cm$^{-3}$) compared to the ambient cloud. In the scenario proposed by \citet{longmore13b}, this cloud has recently passed pericentre with the supermassive black hole at the center of the Galaxy and is being tidally compressed perpendicular to the orbit and stretched along the orbit. Preliminary numerical modeling results suggest the diffuse outer layers of the cloud may be removed in the process, leading to the large observed velocity dispersions and explaining the observed cloud morphology (Kruijssen, Dale \& Longmore in prep.). The bulk of the cloud mass can remain bound, even though standard virial analysis would suggest the cloud is globally unbound \citep[e.g.][]{kauffmann13}. The prediction is that the tidally-injected energy is presently supporting the cloud against collapse but as the cloud continues on its orbit this energy will be dissipated through shocks and the cloud will eventually collapse to form a YMC. Given the large difference between the observed dust temperature \citep{molinari11} and gas temperature \citep{guesten81,mills13}, complicated chemistry, extreme excitation conditions, evidence for widespread shocks and a high cosmic ray rate \citep{ao13,yusefzadeh13}, detailed observational studies (\citet{kendrew13}, \citet{johnston13}, Rathborne et al sub.) and numerical modeling \citep[e.g.][ Kruijssen, Dale, Longmore et al. in prep., Lucas \& Bonnell in prep.]{clark13} are required to test this hypothesis. Understanding the origin and star formation potential of these extreme Galactic Center clouds promises to be an exciting  avenue for study in the next few years.

\subsubsection{YMC formation in the Galactic disk}
\label{subsub:ymc_form_disc}

To date, no clouds of gas mass $\sim$10$^5$\,$\msun$ and radius $\sim$1\,pc with no signs of star formation have been found outside the Galactic Center. So what were the initial conditions for the YMCs that are known to have formed in the disk of the Milky Way? Clues to their origin can be gleaned from the properties of the present-day molecular cloud population in the disk. The mass distribution follows a power law, dN/dM~$\propto$~M$^{-\gamma}$, with $1.5<\gamma<1.8$ \citep{elmegreen96, rosolowsky05}. In order to produce 10$^4$\,$\msun$ of stars, mass conservation means that YMCs must have formed within progenitor clouds of gas mass at least 10$^4$\,$\msun$/$\epsilon$ $\sim$10$^5$\,$\msun$. Clouds more massive than this have virial ratios close to unity (albeit with significant scatter) \citep{rosolowsky07, dobbs11}. While there are many difficulties in using virial ratios to unambiguously determine if an individual cloud is gravitationally bound, it seems reasonable to assume that many of the most massive clouds are likely to be close to being gravitationally bound. This places an interesting constraint -- the fundamental gas reservoir limitation means that the clouds in the disk where one expects to find YMC progenitors are also clouds which are possibly undergoing global gravitational collapse. 

The clouds of gas mass $>10^5\,\msun$ in the disk are typically many tens to hundreds of parsecs in size. The average volume and column density is therefore low (e.g. a few 10$^2$\,cm$^{-3}$, a few 10$^{21}$\,cm$^{-2}$), especially compared to similar mass clouds in the Galactic Center. However, these global size-scales are much larger than the parsec scales of interest for YMC formation. The immediate conclusion is that the YMCs embedded within these clouds can therefore only make up a small volume filling factor of the whole cloud. 

The fact that no 10$^5\,\msun$, pc-scale, starless sub regions have been found within GMCs in the disk suggests that the GMCs do not begin life with such dense subregions. However, we know of at least 18 dense, parsec scale sub regions of $\geq10^4\,\msun$ that have prodigious star formation \citep{ginsburg12}. By learning how these couple to the larger ($10-100$\,pc scale) cloud it may be possible to understand how YMCs assemble their mass in the disk. As mentioned earlier, each of the regions in the first quadrant containing a candidate progenitor cloud has been well-studied and much is known about the gas properties and (embedded) star formation activity. Therefore, such a study is feasible. However, an in-depth review of these detailed studies is beyond the scope of this work. Instead, we focus on two regions: W49 and W43. The former is the most luminous star forming region in the Galaxy and contains the most massive and dense progenitor cloud in the \citet{ginsburg12} sample. As such, this is the most likely site of future YMC formation in the first quadrant. In terms of large-scale Galactic structure, W49 and most other YMC progenitor cloud candidates are not found at any `special' place in the Galaxy (other than potentially lying within spiral arms). W43 is the possible exception to the rule,  and is postulated to lie at the interface between the Scutum-Centaurus (or Scutum-Crux) arm and the stellar bar.

\paragraph{W49:} Lying at a distance of $11.11^{+0.79}_{-0.69}$\,kpc from Earth \citep{zhang13}, W49 is the most luminous star forming region in the Galaxy \citep[10$^{7.2}\,\lsun$:][scaled to the more accurate distance of \citet{zhang13}]{sievers91}, embedded within one of the most massive molecular clouds \citep[$\sim1.1\times10^6\,\msun$ within a radius of 60\,pc;][and references therein]{galvan-madrid13}. The spatial extent of the entire cloud is 120\,pc, but the prominent star formation region, W49A, is confined to an inner radius of $\sim$10\,pc. W49A is comprised of three subregions -- W49N, W49S and W49SW -- each with radii of a few pc and separated from each other by less than 10\,pc. Approximately 20\% of the mass (and practically all of the dense gas) lies within 0.1\% of the volume \citep[$\sim 2\times10^5$\,M$_\odot$ of gas within a radius of 6\,pc:][]{nagy12,galvan-madrid13}. W49N is the most actively star forming of these, containing both a cluster of stars $>4\times10^4\,\msun$ and a well known ring of $\hii$ regions, all within a radius of a few pc \citep{welch87,alves_homeier03,homeier_alves05}. \citet{ginsburg12} identified both W49N and W49SW as likely YMC progenitor clouds. Historically, several scenarios have been put forward to explain the interaction of these dense clumps with the larger scale cloud, from global gravitational collapse to cloud collisions and triggering \citep[e.g.][]{welch87,buckely_wardthompson96}. In the most recent, multi-scale dense gas survey, \citet{galvan-madrid13} show the larger scale gas cloud is constructed of a hierarchical network of filaments that radially converge on to the densest YMC progenitor clouds, which act as a hub for the filaments (reminiscent of the ``hub-filament" formation scenario described in \citet{myers09} and observed towards the very luminous massive star formation region G10.6 by \citet{liubaobab12}). Based on kinematic evidence, they conclude the region as a whole is undergoing gravitational collapse with localized collapse onto the YMC progenitor clouds. 

\paragraph{W43} Located within the so-called `molecular ring', the region lying between Galactic longitudes of 29.5$^\circ$ and 31.5$^\circ$  contains a particularly high concentration of molecular clouds, several well-known star formation complexes such as the mini-starburst W43-Main, and 4 YMC precursors \citep{motte03,ginsburg12}. The $^{13}$CO emission in the region shows complicated velocity structure over 60\,\kms\ along the line of sight. \citet{nguyen_luong11} and \citet{carlhoff13} conclude that almost all of this $^{13}$CO gas is associated with W43, and that the gas within $\sim20\,\kms$ of the $96\,\kms$ velocity component is part of a single W43 molecular cloud complex. This complex has an equivalent diameter of $\sim140$\,pc, total mass of $\sim7\times10^6\,\msun$ and many subregions of high gas density. The measured distance is consistent with the complex lying at the meeting point of the Scutum-Crux arm at the end of bar. \citet{nguyen_luong11} and Motte et al (sub.)  argue that the large velocity dispersion and complicated kinematic structure indicate the convergence point of high velocity gas streams. Three YMC progenitor clouds lie within this longitude-velocity range making it a particularly fertile place for YMC formation. Two of the three progenitor candidates --  W43-MM1 and W43-MM2 -- have projected separations of less than a few pc, and are very likely to be associated. \citet{nguyen_luong13} conclude that these have formed via colliding flows driven by gravity. Given that several red super giant (RSG) clusters are found at a similar location and distance \citep{figer06,davies07}, this region of the Galaxy appears to have been forming dense clusters of $>10^4\,\msun$ for at least 20\,Myr. It also flags the other side of the bar as an interesting place to search for YMCs and YMC progenitor clouds \citep[indeed][have already identified one YMC candidate there]{davies12}.

\subsubsection{Summary: environment matters}

In summary, the disk and Galactic Center are assembling gas into YMCs in different ways. In the Galactic Center, the mechanism is `in-situ slow formation', where the gas is able to reach very high densities without forming stars. Something, possibly cloud-cloud collisions or tidal forces and the gas dissipating energy through shocks, allows some parts of the gas reservoir to collapse under its own gravity to form a YMC. 

Conversely, in the disk, the lack of starless $10^5\,\msun$, r~$<1$pc gas clouds suggest YMCs either form in a `conveyor belt' mode, where stars begin forming as the mass is being accumulated to high density, or the time scale to accumulate the gas to such high densities must be much shorter than the star formation timescale. In two of the most fertile YMC-forming regions in the first quadrant (W49 and W43), recent studies have shown evidence of large-scale gas flows and gravitational collapse feeding the YMC progenitor clouds.

 After YMCs have formed, the remains of their natal clouds can also provide clues to the formation mechanism. Observations of the remaining gas associated with the formation of Westerlund~2 and NGC~3603 suggest these YMCs formed at the interaction zones of cloud-cloud collisions \citep{furukawa09, ohama10, fukui13}. 
 
 From the above evidence alone, it would be premature to claim large scale gas flows as a necessary condition to form YMCs in the disk. However, combined with the fact that the most massive gas clouds have virial ratios closest to unity \citep{rosolowsky07, dobbs11}, and numerous numerical/observational studies of other (generally less massive/dense) cluster forming regions show evidence for large-scale gas flows feeding gas to proto-cluster scales \citep[e.g. W3(OH), G34.3+0.2, G10.6-0.4, SDC335.579-0.292, DR21, K3-50A, Serpens South, GG035.39-00.33, G286.21+0.17, G20.08-0.14~N:][]{keto87a, keto87b, liubaobab12, peretto13, csengeri11, hennemann12, klaassen13, kirk13, henshaw13, nguyen_luong11b, barnes10, smith09,galvan-madrid09}, often with ``hub-filament" morphology \citep{myers09,liubaobab12}, suggests this is a fruitful area for further investigation. 

If YMCs in the disk of the Milky Way form predominantly as a result of large-scale gas flows, one would not expect to see pc-scale regions of $\sim10^5\,\msun$ with no star formation. Rather, the initial conditions of the next YMC generation must be massive GMCs with little signs of current star formation, and kinematic signatures of either large-scale infall or converging flows. Searching for these clouds is another interesting avenue for further investigation.

So what can we learn about YMC formation more generally from this analysis? One potential interpretation of the difference between the YMC progenitor cloud demographics between the first quadrant and inner 200\,pc is evidence for  two `modes' of YMC formation. However, there are several reasons to suspect this may be misleading. Firstly, the physical properties in the interstellar medium are observed to vary continuously, as are the range of environmental conditions in which star formation occurs. Secondly, gas is scale-free. Therefore, despite the fixed mass/radius criterion for YMC progenitor clouds being physically well-motivated in terms of forming bound stellar systems, imposing \emph{any} discrete scale in a scale-free system is arbitrary by definition. The distinction between the simplistic scenarios in $\S$~\ref{sub:simple_formation_scenarios} is therefore sensitive to the criteria used to define likely progenitor clouds, and essentially boils down to whether or not stars have begun forming at the imposed density threshold. 

\begin{figure*}
\includegraphics[height=0.47\textwidth, angle=-90]{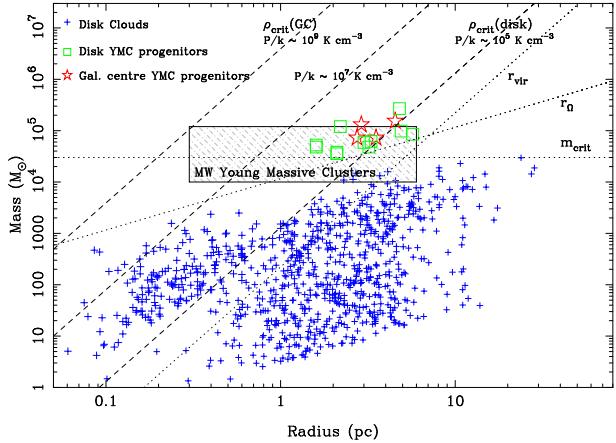}
 \caption{\small  Mass vs radius for gas clouds and YMCs in the Milky Way. Observations of gas clouds are shown as colored symbols and the hatched rectangle shows the location of Galactic YMCs (from \citet{portegieszwart10}). The blue plus symbols are infrared-dark clouds and dense ammonia clouds in the disk \citep{rathborne06, walsh11, purcell12}, the green squares are YMC progenitor candidates in the disk \citep{ginsburg12, urquhart13}, and the red stars are YMC progenitor cloud candidates in the Galactic Center \citep{immer12, longmore12, longmore13b}. The dotted lines show the criteria for YMC progenitor clouds put forward by \citet{bressert12}: m$_{\rm crit} $is the minimum mass reservoir requirement; r$_\Omega$ is the criterium that the escape speed of the cloud is greater than the sound speed in ionized gas ($\sim$10\,$\kms$); r$_{\rm vir} $ is the virial mass given a crossing time of 1Myr. Based on these criteria, clouds lying above all three dotted lines are candidate YMC progenitor clouds. The dashed lines show the critical density for star formation expressed in Eq~\ref{eq:crit_dens} for three different environments. From left to right, the gas properties used to calculate the critical densities are: $v_{\rm turb} = (15, 7, 2)\,\kms$, $\rho_0=(2\times10^4, 10^3, 10^2)$\,cm$^{-3}$, T$= (75, 40, 20)$\,K, with $\alpha_{\rm vir}=1$ in all cases. The corresponding turbulent pressure, expressed as P/k, are given next to each line. The gas properties for the left and right dashed lines are chosen to resemble those for clouds in the Galactic Center and disk, respectively.  }
 \label{fig:mass_radius}
 \end{figure*}

In theories of star formation where turbulence sets the molecular cloud structure \citep[e.g.][]{krumholz05, padoan11, federrath10}, in the absence of magnetic fields, the critical over-density for star formation to begin, $x_{\rm crit}$, is given by $x_{\rm crit} = \rho_c / \rho_0 = \alpha_{\rm vir} {\cal M}^2$, where $\rho_c$ is the critical density for star formation,  $\rho_0$ is the mean density, $\alpha_{\rm vir}$ is the virial parameter and  ${\cal M}$ is the turbulent Mach number of the gas. The critical density is then given by,  $\rho_c~=~\alpha_{\rm vir}\,\rho_0\,{\cal M}^2~=~\alpha_{\rm vir}\,\rho_0\,v_{\rm turb}^2/c_s^2~=~\alpha_{\rm vir}\,P_{\rm turb}/c_s^2$, where $v_{\rm turb}$ is the turbulent velocity dispersion, $c_s$ is the sound speed and $P_{\rm turb}$ is the turbulent pressure.  Therefore, given an imposed mass and radius threshold, M$_{\rm th}$ and R$_{\rm th}$, a molecular cloud in a particular environment will have begun forming stars if it passes the following YMC density threshold, $\rho_{\rm YMC}$, criterion:
\begin{equation}  
\label{eq:crit_dens}
\rho_{\rm YMC} \propto M_{\rm th}/R_{\rm th}^3 > (4\,\pi/3) \, \alpha_{\rm vir}\,P_{\rm turb}\,\mu\,m_H/k_BT
\end{equation}
where $\mu$, $m_H$, $k_B$ and $T$ are the molecular weight, hydrogen mass, Boltzmann constant and gas temperature.
Figure ~\ref{fig:mass_radius} shows the implications of this critical density for gas clouds in different Galactic environments. Clouds with densities below the critical value, but high enough to satisfy the YMC progenitor candidate criteria, form according to the ``in-situ" scenario. Clouds that satisfy the YMC progenitor candidate criteria but with densities exceeding the critical value should already be forming stars, and hence forming YMCs in the ``conveyor belt" mode, or the time scale for gas accumulation is much shorter than the star formation timescale. The YMC candidates in the disk/Galactic Center have average volume densities above/below the critical density for star formation in that environment, respectively. This provides a potential explanation as to why the candidate YMC progenitor clouds in the disk and Galactic Center have similar global properties, but all those in the disk have prodigious star formation activity and many of those in the Galactic Center do not. 

Building on this progress towards a general understanding of YMC formation requires knowing (i) how the dense gas reservoirs got there in the first place [e.g. looking at the kinematics of HI halos surrounding GMCs], and (ii) how similar the Galactic conditions are to other YMC-forming environments across cosmological timescales. 

Numerical simulations and extragalactic observations offer alternative ways to tackle point (i). The review by Dobbs et al. (this volume) shows there are many potential molecular cloud formation mechanisms which become dominant in different environments. All these mechanisms involve flows with some degree of convergence, and it seems plausible that the processes forming YMC progenitor clouds involve those with the largest mass flux to the size scales of interest. Extragalactic observations in starburst systems like the Antennae and M82 show that gas clouds thought to be the sites of future YMC formation lie at locations of colliding gas flows and regions of very high gas compression \citep{keto_ho_lo05, wei12} .

While there appear to be qualitative similarities  across a wide range of environments (e.g. converging gas flows and gravitational collapse in the Milky Way disk and star bursting environments), a more quantitative approach is needed to rigorously compare the environmental conditions and gas properties. Ultimately, the extent to which we can apply what we learn about star and planet formation in our Galaxy to other locations in the Universe depends on how similar the Galactic conditions are to other environments across cosmological timescales. Observational resolution and sensitivity limitations make a direct comparison of regions across such a wide range of distances challenging. However, by taking due care when comparing widely heterogenous datasets, it is possible. In a recent study, \citet{kruijssen13d} show that in terms of their baryonic composition, kinematics, and densities, the clouds in the solar neighborhood are similar to those in nearby galaxies. At the current level of observational precision, the clouds and regions in the Galactic Center are indistinguishable from high-redshift clouds and galaxies. The Milky Way therefore contains large reservoirs of gas with properties directly comparable to most of the known range of star formation environments and is therefore an excellent template for studying star and planet formation across cosmological time-scales. 

Returning to the focus of the review, how does what we are learning about YMC formation affect the role of YMCs as a probe of star and planet formation in extreme environments? Irrespective of the details, the above analysis suggests that clusters of the same final mass and radius can have very different formation histories. In which case, if investigating the role of formation environment on the resulting stellar/planetary population in a YMC, it is crucial to understand the cluster's formation history. Both in terms of the initial (proto)stellar densities and gas conditions, the YMCs forming in the Galactic Center represent the most extreme conditions for star/planet formation in the Galaxy. This strongly motivates comparing the detailed star formation process in the Galactic Center with that in the rest of the Galaxy. The fact that to date, key parameters (like the IMF) of YMCs in the Galactic Center clusters and those in the disk are consistent to the uncertainty level of current observations  \citep[e.g.][although the nuclear cluster may be an exception \citep{jessicalu13}]{bastian10,habibi13,hussmann12} suggests that the extreme environment and high protostellar densities do not affect how stars assemble their mass, at least in a statistical sense. Finally, the fact that the range of star forming environments in the Milky Way are so similar to the range of known star forming environments across cosmological timescales suggests one can directly apply what we learn about star and planet formation in different parts of the Galaxy to much of the Universe.

\section{\textbf{The Early Evolution of Young Massive Clusters}}

This section is devoted to the first $\sim$10 Myr of a YMC's lifetime.  Over
the past few years we have seen substantial progress in our
understanding of this evolutionary phase, both observationally and
theoretically. In two subsections, we first review the current
theoretical framework (\S~3.1) before building on this to summarize the
intense recent debate on two key open questions in the field (\S~3.2).

\bigskip
\noindent

\subsection{\textbf{Theoretical framework}}
\label{sub:theory}

The formation and early evolution of YMCs is dominated by a plethora of complex and interacting physical processes. A theoretical understanding of this phase requires insight in the conversion of gas into stars \citep{elmegreen00,clarke00,krumholz07}, in the feedback effects of individual stars on the cloud as a whole and locally \citep{dale05,murray10,dale12,krumholz12}, as well as the (possibly differing) phase-space distribution of the gas and the stars -- both before and after feedback effects become important \citep{offner09b,moeckel10,kruijssen12,girichidis12}. The interplay of these processes determines the star formation efficiencies (SFEs) in the proto-YMC clouds, the timescale on which gas not involved in star formation is expelled, and therefore, ultimately, what fraction of the stars reside in bound clusters once star formation is complete \citep{kruijssen12d}. Once all the gas in such clouds has been converted to stars or expelled by feedback, the modelling of a YMC reduces to solving the gravitational $N$-body problem. Even in the absence of external forces, the evolution of such a dense stellar system is very complex, involving processes such as mass segregation, core collapse, stellar collisions and the dynamical evaporation of the YMC \citep{portegieszwart10}. In addition, YMCs can be affected by close interactions with molecular clouds or other density peaks in their galactic environment \citep{spitzer87,gieles06,kruijssen11}. In this section, we review the new generation of hydrodynamical and $N$-body simulations which are able to model these processes and their effects on star formation in massive, dense molecular clouds.

Despite the recent advances in the numerical modelling of cluster formation (see \citealt{kruijssen13} for a review), there are very few simulations of YMC formation, mainly due to the large dynamic range that needs to be covered in terms of mass, size, and time. Current state-of-the art simulations can model the formation of a cluster with a stellar mass spectrum that is complete down to the hydrogen burning limit ($\sim0.08~\msun$) up to mass scales of a few $10^3~\msun$ \citep{tilley07,bonnell08,krumholz12}, still well below the lower mass limit of the YMC regime. As a result, any current simulation aiming to model the formation and early evolution of YMCs necessarily has to neglect possibly important physics, such as resolving the assembly of the full stellar mass spectrum. A second problem is that simulating feedback processes (e.g. ionization, stellar winds, supernovae) is computationally expensive for several reasons. For instance, a full, three-dimensional modelling of radiative transfer is intrinsically very expensive. Additionally, feedback effects exacerbate the problem of the dynamic range that needs to be covered, in terms of mass, time and spatial scales (gas ejected by stellar winds or supernovae can reach velocities of several $1000~\kms$, while cold gas has a typical velocity dispersion of $1$--$10~\kms$).

Because of the above limitations, it is necessary that current and future studies addressing different parts of the YMC formation problem are well-connected, self-consistently linking the different physics. We therefore review the theoretical framework for the formation and early evolution of YMCs with a view to setting out the initial conditions from which such simulations should begin. This framework is divided in three steps that are often probed in different numerical experiments. Linking these steps will be one of the major challenges of the coming years.

\subsubsection{YMC formation from the hierarchical ISM}
A theory of YMC formation (or cluster formation in general) first needs to acknowledge the hierarchical nature of the ISM \citep{elmegreen96,klessen98,bate98,padoan02}, which persists in the condensation and assembly of GMCs and protostellar cores, reaching all the way from the galactic gas disk scale height ($\sim100~\pc$) to the ambipolar diffusion length ($\sim0.01~\pc$). The term {\it hierarchy} is often used very loosely to indicate scale-free or fractal-like substructure -- for clarity, we adopt the definition that hierarchicality indicates the growth of structure from the merging of smaller structures. Hierarchical growth is crucial for understanding YMC formation, as even the most extreme and densest proto-cluster clouds like the Brick \citep{longmore12,rathborne13} show substructure. If star formation is driven by self-gravity, then the hierarchical structure of gas clouds implies that the collapse to protostellar cores occurs more rapidly locally than on a global scale, and hence the conversion to the near-spherical symmetry that characterizes YMCs has to take place when protostellar cores and/or stars have already formed. The hierarchical growth of proto-YMC clouds therefore continues during star formation and the assembly of the actual YMC, during which the gravitational pull of the global gas structure may provide a `conveyor belt' for transporting stellar aggregates to the center of mass.

Numerical simulations that include self-gravity, hydrodynamics, and sink particle formation automatically capture the hierarchical growth of YMCs, provided that they start from turbulent initial conditions \citep[see e.g.][]{maclow04} or convergent flows. The turbulence generates scale-free structure in which perturbations on the smallest length-scales naturally collapse first, leading to star formation dispersed throughout much of the cloud volume. The work by \citet[]{maschberger10} (see their Figure 3) illustrates how the resulting merger trees of the stellar structure in numerical simulations of cluster formation can be used to trace and quantify hierarchical growth -- much like their use in galaxy formation studies.

\subsubsection{The early disruption of stellar structure}
While the global dynamics of gas and stars are initially coupled, there are several reasons why their co-evolution cannot persist for longer than a dynamical time. Firstly, gas is dissipational whereas the stars are ballistic, which inevitably leads to differing kinematics. Secondly, gas is being used to form stars -- because gravitational collapse and star formation proceed most rapidly near the density peaks, these attain much higher local SFEs than their surroundings, possibly appearing as relatively gas-free, gravitationally bound stellar groups within a cocoon of gas \citep{peters10,kruijssen12,girichidis12}. In numerical simulations, these stellar groups are virialised even when omitting the gravitational influence of the gas. The local gas exhaustion is driven by accretion as well as the accretion-induced shrinkage of the stellar structure (while gas expulsion leads to the expansion of an initially virialised system, gas accretion has the opposite effect), implying that the gas and stars are indeed decoupled \citep[as is also indicated by the velocity dispersions of the gas and stars, see][]{offner09b}.

If the gas density is not high enough to deplete the gas through secular star formation processes, the co-evolution of gas and stars ceases when stellar feedback becomes important. Protostellar outflows \citep{nakamura07,wang10,hansen12,krumholz12}, photoionising radiation \citep{dale12}, radiative feedback \citep{offner09,murray10}, stellar winds and supernovae \citep{pelupessy12} can potentially blow out large mass fractions. Whether or not the expulsion of residual gas by feedback affects the boundedness of stellar structure depends on the division between gas exhaustion and gas expulsion. If the density was high enough to lead to gas exhaustion, then gas expulsion cannot affect the dynamical state of the stars, whereas a low SFE (and hence density) implies that the stars themselves are only held together by the gravitational potential of the gas, and therefore will disperse when the gas is expelled. The obvious question to ask is whether such low-density stellar associations were gravitationally bound at any point during the star formation process. In part, this depends on the dynamical state of their nascent clouds. Current evidence suggests that at least some substantial fraction of clouds and star-forming regions is initially globally unbound \citep[e.g.][]{heyer09,dobbs11}. Simulations of globally-unbound clouds by \citet{clark05} show that the clusters formed in such an environment resemble OB--associations containing several locally-bound but mutually unbound subgroups.
 
It is clear that the importance of feedback and gas expulsion crucially depends on the outcome of the star formation process. Dense, virialised YMCs likely experienced a formation history that was dominated by gas {\it exhaustion}, whereas unbound associations suffered from the effects of gas {\it expulsion}. The balance between both mechanisms can vary among different parts of the same region, or when considering different spatial scales. It seems logical that there exists a critical density, irrespective of the spatial scale, above which gas exhaustion is more important than gas expulsion. A second, independent mechanism for the early disruption of stellar structure is formed by the tidal perturbations coming from the dense, star-forming environment (the `cruel cradle effect', \citealt{kruijssen11}; also see \citealt{elmegreen10b}), which is thought to suppress the formation of bound clusters at galactic gas surface densities of $\Sigma>10^3~{\rm M}_\odot~{\rm pc}^{-2}$ \citep{kruijssen12d} and primarily affects low-density clusters. The key conclusion is that, irrespective of the particular mechanism, the early disruption of stellar structure does not depend on mass or spatial scale, but exclusively on the volume density.

For a long time, our picture of the early evolution of stellar clusters was based on numerical simulations lacking a live gas component, i.e.,~$N$-body simulations of stars existing in virial equilibrium with a background potential representing the gas -- the latter being removed to probe the effects of gas expulsion \citep{lada84,goodwin97,adams00,geyer01,boily03,goodwin06,baumgardt07}. These simulations naturally showed that if a large part of the gravitational potential is removed, a corresponding part of the structure is unbound \citep[as can be shown analytically, see][]{hills80}. Together with the observation that after some 10~Myr only 10\% of all stars are situated in bound clusters \citep{lada03}, this has led to the classical picture of early cluster evolution, in which 90\% of all clusters are destroyed on a short time-scale by gas expulsion. By contrast, the modern view of cluster formation discussed above connects gas expulsion to unbound associations, standing separately from the early evolution of the gravitationally bound, dense clusters (of which YMCs are the most extreme examples) that represent the high-density end of a continuous density spectrum of star formation. This separation implies a prediction that the fraction of star formation occurring in bound clusters varies with the galactic environment \citep{kruijssen12d}, and it crucially depends on the spatial variation of the SFE as a function of the local volume density and the resulting relative spatial distributions of gas and stars. It is therefore key to develop physically-motivated initial conditions for future numerical simulations of early cluster evolution that are designed to accurately represent these distributions.

\subsubsection{When do stellar dynamics take over?}
In dynamically young systems, the hierarchical growth of gaseous structure translates to a hierarchy in the young stellar structure. However, in gravitationally bound systems (i.e.,~those with ages older than a dynamical time-scale, see \citealt{gieles11}), this hierarchy is short-lived -- once the stellar mass dominates self-gravity on a certain spatial scale (starting at the smallest scales), the hierarchy is erased by $N$-body dynamics on a dynamical time-scale. The step towards stellar dynamics-dominated evolution may occur due to gas exhaustion at high densities, gas expulsion at low densities, or a combination of both on large spatial scales covering a wide range of local environments.

Once the gravitational support from the gas has vanished, the hierarchical growth of stellar structure may continue if the region is globally sub-virial, but only for another dynamical time-scale \citep{allison10}. The merging of sub-clusters leads to violent relaxation and the growth of a centrally concentrated stellar system, during which the degree of mass segregation in the progenitor systems is preserved \citep{allison09}. As a result, mass segregation can be achieved on a dynamical time-scale, even in massive YMCs. The further structural evolution of a cluster or YMC is governed by two-body relaxation and possible external tidal perturbations.

\subsubsection{Towards physically motivated initial conditions for numerical simulations}
We have divided the cluster formation process into three different stages:
\begin{itemize}
\item[(1)]
The condensation of stars and stellar structure from the hierarchically structured ISM.
\item[(2)]
The early disruption of stellar structure by gas expulsion or tidal perturbations. By contrast, violent relaxation erases sub-structure but does enable the formation of YMCs and long-lived clusters.
\item[(3)]
The transition to the phase where YMC evolution is dominated by stellar dynamics.
\end{itemize}
Each of these stages have been studied separately in numerical simulations of cluster formation, using a variety of numerical methods. It is possible to base the initial conditions for each of these steps on the outcomes of the previous ones. For instance, the initial conditions for YMC formation from the turbulent, hierarchical ISM requires one to resimulate (i.e.,~zoom in on) GMCs identified in galactic-scale simulations \citep[e.g.][]{tasker09,dobbs11b}. The next step demands the inclusion of realistic feedback and star formation models -- as reviewed in \citet{kruijssen13}, such simulations should ideally include radiative feedback, protostellar outflows, and magnetic fields. Using the hierarchically structured initial conditions from the preceding step, and ensuring these conditions accurately reflect the observed properties of such regions, will ensure a physically accurate setup. 

The transition from an embedded cluster to a gas-poor system is non-trivial and benefits from a fully self-consistent treatment of both the collisional stellar dynamics as well as the hydrodynamics. This has been enabled by a new generation of numerical methods such as {\sc Amuse} \citep{portegieszwart09} and {\sc Seren} \citep{hubber13}. If the classical approach of integrating the $N$-body dynamics in a background potential representing the gas is to remain viable, it is key that the gravitational potential is taken from hydrodynamical simulations \citep[e.g.][]{moeckel10}. These should either be seeded with stars at the density peaks, or the sink particle data from the prior simulation could be used for the initial conditions \citep[e.g.][]{parker13}. While this method is intrinsically flawed with respect to more advanced numerical approaches that self-consistently solve for sink particle formation, feedback, and the co-evolution of gas and stars, it can still lead to new insights if it is applied with care.

In the near future, simulations of YMC formation are facing two main challenges. The first lies in establishing a consensus on scale-free (or generalized) initial conditions for any of the above three stages. The above discussion provides a physically motivated outline for how this can be done. The second challenge is one of physical scale -- a YMC by definition has a stellar mass $M>10^4~{\rm M}_\odot$, which has not successfully been reproduced in numerical simulations without sacrificing the accuracy of low-mass star formation and the resulting stellar dynamics \citep{tilley07,krumholz12,dale12}. As long as computational facilities cannot yet support the fully self-consistent modelling of YMC formation, such physical compromises may even be desirable. After all, answering problems such as the appearance of multiple generations of stars in massive clusters \citep{gratton12} or the formation of massive black holes \citep{portegieszwart04} likely require the numerical modelling of YMC formation. Especially in the case of massive YMCs, the omission of accurate stellar dynamics will not necessarily be unsurmountable -- the relaxation times of YMCs exceed the time-scales of interest when modelling their formation, and the first large-scale numerical simulations of spatially resolved YMC formation in a galactic context \citep{hopkins13} provide an excellent benchmark for future, more detailed efforts.


\subsection{Key Open Questions}

\bigskip
\noindent
\subsubsection{The distribution of young stars}
\label{sub sub:dist_young_stars}
\bigskip

How long do clusters live and what causes their destruction?  This appears to be a straight forward question -- surely one can simply look at a population of young clusters and look for a characteristic timescale at which they begin to disappear?  In reality, this is a nuanced problem. On the one hand, exposed (i.e., non-embedded), dynamically evolved clusters are distinct objects. However, young clusters often contain smaller sub-structures with no distinct boundaries  within much larger ensembles (c.f., \citealt{allen07}).  This is presumably due to the scale-free nature of the gas out of which clusters form (e.g., \citealt{hopkins13b}).  

In their seminal work, \citet{lada03} studied a collection of embedded young regions, and defined clusters to be groups with $>35$ members with a stellar mass density of $>1\msun$\,pc$^{-3}$, which translates to a stellar surface density of 3 YSOs/pc$^2$.  With this definition, they determined that $\sim90$\% of stars formed in clusters.  When comparing this value to the number of open clusters (relative to the number expected from embedded clusters) they concluded that the majority of clusters disrupted when passing from the embedded to exposed phase.  One caveat to this work is that the open cluster catalogue used by \citet{lada03} was only complete to a few hundred parsecs, whereas they assumed that it was complete to 2~kpc.  This does not strongly affect their results for the early evolution, but it does explain why their expected open cluster numbers continue to diverge when compared to observations of clusters older than 5-10~Myr.

\citet{bressert10} used a comprehensive Spitzer Space Telescope Survey of nearly all YSOs in the solar neighborhood (within a distance of 500~pc) to measure their surface density distribution.  They found a continuous distribution (i.e. without a characteristic density), with a peak at $\sim20$~YSOs/pc$^2$, log-normally distributed (potentially with an extended tail at high densities).  The authors concluded that the fraction of stars forming in clusters is a sensitive function of the criteria used to define a cluster.  By adopting different criteria proposed in the literature, one could conclude that between 20\% to 90\% of stars form in clusters.  The ambiguity in defining a cluster makes it difficult to quantify a single mechanism or efficiency that can explain the disruption of clusters across the full range of mass and density.

The initial spatial distribution of stars plays an important role in the subsequent evolution of the cluster. Both a centrally concentrated cluster profile (e.g., \citealt{bate98}; \citealt{parker12b}; \citealt{pfalzner12}), and a rapid expansion of initially dense low-N clusters due to dynamical relaxation (e.g., \citealt{gieles12}) can reproduce the continuous surface density distribution observed by \citet{bressert10}. However, the observed spatial correlation of many YSOs with filamentary gas (c.f., \citealt{allen07}; \citealt{gutermuth11}), combined with the lack of strong deviations between YSOs in various phases, argues against this set of initial conditions.

The continuous distribution of structures as a function of spatial scale makes it difficult to define what constitutes a (bound) cluster or an (unbound) association.  While it is clear that stars rarely form in isolation, it does not appear to be the case that all stars form in gravitationally-bound clusters.  The terminology used for these young regions (e.g., clusters, groupings, ``clusterings", associations) has led to much confusion and misunderstanding within the field.

In order to better characterize the different stellar structures found in SF regions, \citet{gieles11} introduced a dynamical definition of clusters/associations.  The authors focus on massive stellar populations, $>10^4$\,\msun, and found multiple examples where the same region was referred to as a cluster and an association by different studies.  To address this problem, they introduced the parameter $\Pi$ -- the ratio between the stellar age and the crossing time of a region.  Since unbound regions expand, the crossing time continually increases, so $\Pi$ stays roughly constant or decreases with time.  By comparison, the crossing time for bound structures remains fairly constant, so $\Pi$ values increase with age.  With this definition, \citet{gieles11} looked at a collection of regions discussed in the literature.  They found that at young ages ($<3$~Myr) there was a continuous distribution of $\Pi$ values.  After 10~Myr, the distribution became discontinuous with a break at $\Pi=1$, clearly separating expanding associations ($\Pi < 1$) and bound clusters ($\Pi > 1$).  

To summarize, since the gas distribution from which stars form is scale-free, it is not expected that all stars form in dense or gravitationally-bound clusters. Rather, young stellar structures are expected to have a continuous distribution of densities \citep{hopkins13b}.  Hence, at young ages it is non-trivial to define which stars will end up in bound clusters.  It is only once a structure has dynamically evolved (i.e., is older than a few crossing times) that structures become distinct from their surroundings.

While it may not be possible to define the fraction of stars that are born in clusters, we can measure the fraction of stars that are in dynamically-evolved ($>$10\,Myr) clusters at a given age.  A variety of studies have found fractions of $\sim10$\% for the Milky Way (e.g., \citealt{miller78}; \citealt{lamers06a}), for ages between a few Myr and a few hundred Myr.  While limited by resolution, extragalactic studies have found fractions between $2-30$\% \citep{adamo11}. In addition, evidence has been found for a potential relationship between the fraction of stars forming within clusters and the star-formation rate surface density (i.e. clusters are more prevalent in starbursts; e.g. \citealt{adamo11,silvavilla_larsen11}).  The techniques used for extragalactic studies are still developing, providing a rich potential avenue for future works in this area.

\bigskip
\noindent
\subsubsection{Infant (im)mortality}
\bigskip

In the simplistic view of cluster formation discussed in \S~\ref{sub:theory}, whether or not a cluster survives the transition from the embedded to exposed phase is related to the effective star-formation efficiency (eSFE -- a measure of how
far out of virial equilibrium the cluster is after gas expulsion, see e.g. \citealt{goodwin09}).  In this picture, if the eSFE is higher than a critical value ($\sim30$\%), a cluster is expected to remain bound.  The destruction of clusters based on the removal of their natal gas (gas expulsion) is generally referred to as ``infant mortality". This is often thought to be mass independent due to the similarity between the embedded and exposed cluster mass functions.  

For massive clusters, the rapid removal of gas and the subsequent expansion of the cluster (due to the resulting shallower gravitational potential), is predicted to have a measurable effect on the clusters' dynamical stability.  For example, \citet{goodwin06} ran a series of N-body simulations of massive clusters undergoing rapid gas removal. They found that the light-to-mass ratio (L/M) varied by more than a factor of 10 for eSFEs of $10-60$\%.  Measurements of the L/M ratio of extragalactic YMCs appeared to confirm that many were expanding. However, \citet{gieles10c} showed that massive binaries are likely affecting the results.

In order to overcome the problem of unresolved binaries, a number of works have focussed on YMCs in the Galaxy and LMC. In these systems it is possible to determine velocities of individual stars either through proper motions or radial velocities, and multiple-epoch velocity measurements can remove binaries from the sample.  So far, this has been done for NGC~3603 \citep{rochau10}, Westerlund 1 (\citealt{mengel07}; \citealt{cottaar12}), the Arches \citep{clarkson12} and R136 in the LMC \citep{henault12}. All clusters studied to date are consistent with being in virial equilibrium, and do not show evidence for being affected by gas expulsion. YMCs are expected to revirialise within 20$-$50 crossing times \citep[$>$4\,Myr for typical observed initial sizes of $\sim$1\,pc and velocity dispersions of $\sim$5\,$\kms$:][]{baumgardt07, goodwin06}, so we would expect to see evidence of this  in the above studies. Therefore, it appears that YMCs are dynamically stable from a very young age \citep[the Arches is $2.5-4$~Myr old; ][]{martins08}. \S~\ref{sub:theory} discussed possible interpretations of this result.

The apparent long term survival of clusters (after the gas expulsion process) is also seen in the age distribution of clusters in the Milky Way (e.g., \citealt{wielen71}).  For example, using complete open cluster catalogues, \citet{lamers05} and \citet{piskunov06} found a flat age distribution (number of clusters per Myr) until $>300$~Myr in the solar neighborhood (600 and 800~pc, respectively).  However, the estimated disruption timescale for these samples are significantly less than that found for the LMC and SMC (e.g., \citealt{hodge87}; \citet{portegieszwart10}). 

The cause of this cluster disruption is likely to be the combined effects of tidal fields, dynamical relaxation, and most importantly for young clusters, interactions between clusters and giant molecular clouds \citep{gieles06,kruijssen11}.  Hence, it is the gas surface density that likely sets the average lifetimes of YMCs in galaxies (c.f., \citealt{portegieszwart10}).

\bigskip
\noindent
\subsubsection{Age Spreads within YMCs}
\label{subsub:age_spreads}
\bigskip

It is often assumed that all stars within a massive cluster are coeval, but how close is this approximation to reality?  If star-formation is hierarchical (e.g., \citealt{efremov98}; \citealt{hopkins13b}), then we would expect the age spread within any region to be proportional to the size of the region studied.  Hence, regions of $20-50$~pc across may have spreads of $\sim10$~Myr, whereas dense compact regions of cluster sizes ($1-3$~pc) are expected to have spreads $<1$~Myr (e.g. \citealt{efremov98}). Here we focus on clusters scales, i.e. $\lesssim3$~pc.  However, we note that young clusters are generally not found in isolation, and that when one looks at the surrounding distribution, age spreads are often found \citep[again, proportional to the size scale probed: see e.g.][]{melena08,roman-zuniga08,dario10a,preibisch11}.

As a general rule, the current observational and theoretical limitations mean the detectable spread in stellar ages within a cluster is proportional to the mean age of that system. Observations of the youngest clusters therefore provide the best targets to quantify absolute age spreads. There are two complimentary ways to determine the ages of young stellar systems (see Soderblom et al this volume for a review on this topic). The first is through spectroscopy of the high mass stars, whose evolutionary timescale is quite short, to place them on the Hertzsprung-Russell diagram. The second method is through photometry of many low-mass pre-main sequence stars in order to place them in a color-magnitude diagram (CMD) and compare the mean age and scatter relative to theoretical models. As an aside, it is worth noting that in addition to being a fundamental question in cluster formation, potential age spreads must be taken into account for accurate derivations of the stellar initial mass function (see Offner et al this volume).  

One potential caveat to studies of young clusters is that clusters tend not to form in isolation but rather within larger star-forming complexes, often containing multiple clusters and a distributed network of stars.  These complexes often have age spreads of the order of 10~Myr, which means that simply due to projection effects, some relatively old (10~Myr) stars may appear to be part of a cluster.  Hence, any measured spread is always an upper limit.  An example of such a region is the Carina Nebula complex, which extends for $\sim30$~pc and contains compact clusters with ages between $\sim1$ and $\sim8$~Myr, along with a large ($\sim50$\% of the total) distributed stellar population (e.g., \citealt{preibisch11,ascenso07}).

The nearest, and best studied cluster, with a mass in excess of 1000~\msun\ is the ONC at a distance of 417$\pm7$\,pc \citep{menten07}.  There has been a long standing debate on the absolute age and relative age spread within the cluster, with initial reports of an accelerating star-formation history (SFH) over the past $\sim10$~Myr, peaking $1-2$~Myr ago \citep{palla_stahler99}.  \citet{dario10} found an age spread of $\sim2-3$~Myr in the cluster by studying the pre-main-sequence (PMS) color distributions with a collection of archival HST images.  However, \citet{alves12} have shown that the cluster/association NGC~1980, which has an older age ($4-5$~Myr),  overlaps significantly in projection, explaining much of the previously reported spread.  Additionally, rotation in young stars can significantly add the dispersions in CMDs of PMSs, which will lead to an over-estimation of any age spread present \citep{jeffries07}.

The most massive YMC known in the galaxy, Westerlund~1 has a mean age of $\sim5$~Myr.  \citet{kudryavtseva12} have studied the PMS color distribution and reported an upper limit to the relative age spread of 0.4~Myr.  It appears that the statistic used did not show the age spread, but rather this is the potential error on the best fitting age.  However, it is clear that the age spread is much smaller than the age of the cluster.  A similar conclusion was reached for the massive cluster R136 in the LMC, based on spectroscopy of the high mass stars (e.g., \citealt{massey98}).  There is also evidence of a $1-2$~Myr age spread within R136, potentially due to the merger of two clusters with slightly different ages \citep{sabbi12}.

The Galactic YMC NGC~3603 has been subject to significant debate on the potential age spread within clusters.  Using the early release science data from WFC3 on HST, \citet{beccari10} reported that $\sim2/3$ of the PMS stars have ages $\sim3$~Myr, while $\sim1/3$ have ages $>10$~Myr (from $0.2$ to $2$~pc from the cluster center).  However, while the spatial distribution of the young stars was centrally concentrated on the cluster, the older stars appear to have a uniform distribution across the field of view.  NGC~3603 is part of a large star forming region, with ongoing star formation. The presence of the evolved massive star Sher~25 has been used as evidence for  a significant duration of star-formation within the region ($>$10\,Myr), not necessarily confined to the cluster itself. However, \citet{melena08} find the age of Sher~25 to be 4\,Myr and subsequently revise the upper age limit in NGC 3603 from 10 to 4 Myr. \citet{kudryavtseva12} also studied the inner 0.5~pc of NGC~3603 and found an upper limit to the age spread of 0.1\,Myr (although see note above).  Hence, this cluster is consistent with a near coeval population, with an older population of stars in the region around it.

Contrary to the above studies, which are consistent with age spreads much less than the mean stellar age, the young Galactic cluster W3-main ($\sim 4 \times 10^3$\,\msun) appears to have an age spread of $\sim2-3$~Myr, similar to the age of the cluster.  This is based on spectroscopy of a massive star that appears to have already left the zero-age main sequence along with the presence of multiple ultra-compact H{\sc ii} regions, all within a projected radius of less than 2\,pc \citep{bik12}. Given the proximity of other nearby star forming complexes, projection effects may explain the observations. However, \citet{bik12} propose that the spatial distribution of the high mass stars and ultra-compact H{\sc ii} regions argues against this. The cluster also appears to have a significant amount of dense gas within the cluster region, potentially explaining why star formation has been able to continue.  

There has been increasing evidence that high-mass and intermediate-mass stars cannot be explained by evolutionary models with the same age. Such an age discrepancy was discussed for Westerlund~1 in \citet{negueruela10}, where red supergiants are expected to be of apparently older age ($> 6$ Myr), while the rich WR population should be younger than 5 Myr (see also \citealt{crowther06}), which is consistent
with the derived age of blue supergiants. A similar trend is found in the Arches and Quintuplet Galactic Center clusters, where the apparent age of intermediate-mass, less evolved supergiants appears to exceed the age of the WN population (\citealt{martins08}; \citealt{liermann10}; \citealt{liermann12}). In all these clusters, particularly young ages are derived from single-star evolution models for the hydrogen-rich WN population. While these discrepancies have led to claims of age spreads in young star clusters, there is increasing evidence that the super-luminous WNs (and LBVs) are rejuvenated binary products (see especially the discussion in \citealt{liermann12}). Hence, while age spreads in the range of $\pm 2$ Myr (e.g., W3 Main, \citealt{bik12}) are present in young associations, there is little evidence for extreme age spreads in the spatially confined population of YMCs (see above references), especially when the expected consequences of binary evolution are taken into account.

\bigskip
\noindent
\textbf{Timescale of the removal of gas from YMCs}
\bigskip

Presumably, the lack of significant age spreads within most YMCs is due to the fact that the dense gas, which is required for star-formation, is removed on short timescales.  For example, YMCs such as the Arches (2.5$-$4~Myr), NGC~3603 (1$-$2~Myr), Trumpler 14 ($\sim1$~Myr) and Westerlund 1 ($5-7$~Myr) are all devoid of dense gas.  It is often assumed that supernovae play a role in the removal of the gas, but as the above examples show, the gas is removed well before the first SNe can occur ($\sim3$~Myr).  Even for relatively low-mass clusters ($\lesssim10^3$\msun) the timescale of gas removal is $<1$~Myr \citep{seale12}.  Potential causes of the rapid gas removal are discussed in \S~\ref{sub:theory}.

This rapid removal naturally explains the lack of significant age spreads in clusters, but places strong constraints on the dynamical stability of young clusters.  

\bigskip
\noindent
\subsection{Age Spreads within Older Clusters}
\bigskip

The ability to discern age spreads within clusters is linked to the (mean) age of the cluster itself, with an uncertainty of approximately 10-20\% of the cluster age.  While the constraints from older clusters are significantly less in an absolute sense, they are still worth looking into as some surprising features have been found.

The relatively nearby double cluster, h \& $\chi$-Persei have masses of 4.7 and $3.7 \times 10^3$\msun, respectively, within the clusters, and a total mass of $2 \times 10^4$\msun, including the common halo around the clusters \citep{currie10}.  Their common age is derived to be 14\,Myr \citep{currie10}. An upper limit of $2$~Myr for the possible age spread has been put on this population using an extensive spectroscopic survey of cluster members \citep{currie10}.

There have been claims of extended star-formation episodes ($200-500$~Myr) within 1-2~Gyr clusters in the LMC/SMC (e.g., \citealt{mackey07}; \citealt{goudfrooij11}) based primarily on the presence of extended main sequence turn-offs in these clusters.  \citet{goudfrooij11} and \citet{conroy11} suggest that such age spreads can happen in massive clusters, which have high enough escape velocities to retain stellar ejecta and accreted material from the surroundings.  \citet{bastian13a} have tested this claim by studying two massive ($10^5$\msun) clusters in the LMC with ages of 180 and 280~Myr, with resolved HST photometry.  The authors do not find any evidence for an age spread within these clusters, and put an upper limit of 35~Myr on any potential age spread.

\citet{bastian13} have compiled a list of all known Galactic and extragalactic massive clusters with ages between $10-1000$~Myr and masses between $10^4 - 10^8$\msun with published integrated spectroscopy and/or resolved photometry of individual stars.  The authors looked for emission lines associated with the clusters, in order to find evidence for ongoing star-formation within the existing clusters.  Of the 129 clusters in their sample, no clusters were found with ongoing star-formation.  This effectively rules out theories and interpretations that invoke extended (tens to hundreds of Myr) star-formation periods within massive clusters (e.g., \citealt{conroy11}).

We will return to the possibility of multiple star-formation events within massive clusters in \S~5, when we discuss globular clusters and how they relate to YMCs.


\section{\textbf{The Effect of Environment on Planet Formation and the Origin of the Field}}

Young stars in YMCs show infrared excess emission and evidence for accretion \citep[e.g.][for the Galactic YMC NGC~3603]{stolte04,beccari10} suggesting the presence of circumstellar disks. Whether these disks are similar to those in nearby environments \citep[see][]{stolte10} or can survive and produce -- possibly even habitable -- planetary systems in YMCs remain interesting questions. Considering the disruptive effects of stellar clustering on the survival of protoplanetary disks and planetary systems in small-$N$ systems \citep[][Rosotti et al. 2013]{scally01,olczak10,parker12,dejuanovelar12}, the formation of (habitable) planets in YMCs may still be possible, but their chances of long-term ($>10^8$~yr) survival are most likely very limited. This is underlined by the absence of planets in the 10~Gyr-old globular clusters NGC~104 and NGC~6397 \citep{gilliland00,nascimbeni12}, whereas Neptune-like planets have been detected in the intermediate age ($\tau\sim1$~Gyr) open cluster NGC~6811 \citep{meibom13}, which has a much lower stellar density ($\Sigma\sim10^3~{\rm pc}^{-2}$) than globular clusters or YMCs ($\Sigma>10^4~{\rm pc}^{-2}$).

Provided that long-term habitable planets cannot form in YMCs, we can now ask the question, what fraction of all stars may be born in such a hostile environment? It was proposed by \citet{dejuanovelar12} that a stellar surface density of at most $\Sigma<10^3~\msun~{\rm pc}^{-2}$ is needed to keep the habitable zone of G-dwarfs unperturbed over Gyr time-scales, whereas \citet{thompson13} showed that protoplanetary systems with densities $\Sigma>6\times10^3~\msun~{\rm pc}^{-2}$ may not have an ice line, inhibiting the formation of gas and ice giants (if they form by core accretion) as well as potentially habitable planets. In the following, we adopt a maximum density for the long-term existence of habitable planets of $\Sigma_{\rm crit}=10^3~\msun~{\rm pc}^{-2}$. Given the very weak mass-radius relation of young clusters $R=3.75~{\rm pc}\,(M/10^4~\msun)^{0.1}$ from \citet{larsen04b}, this can be converted to a critical cluster mass of $M_{\rm crit}\sim10^5~\msun$.

We point out here that a seemingly opposite conclusion was reached by \citet{dukes12} and \citet{craig13}, who find that the dense cluster environment does not affect the properties of disks or planetary systems. However, this apparent dichotomy arises from the adopted definition of `cluster'.  In this review we use the term `cluster' to refer to bound systems, whereas in the above two papers the term refers to any overdensity of stars, reflected by the short lifetimes (of a few crossing times) and constant surface densities (as appropriate for GMCs and unbound associations). This results in a decreasing volume density with mass and hence a decreasing importance of environmental effects. The low-volume density systems considered by \citet{dukes12} and \citet{craig13} constitute the majority of star-forming environments.

The initial cluster mass function is a power law ${\rm d}N/{\rm d}M\propto M^{-2}$ for masses $M>10^2~\msun$ up to some maximum mass \citep{portegieszwart10}, which in the nearby Universe is $M\sim2\times10^5~\msun$ \citep{larsen09}, and in the high-redshift Universe may have been $M\sim10^7~\msun$ (see \S\ref{sec:glob_clusters}). Because the slope of the initial cluster mass function is $-2$, equal numbers of stars are formed in each dex of cluster mass. Given the above mass limits, in the nearby Universe therefore about 10\% of all gravitationally bound clusters are hostile to habitable planets, whereas in the high-redshift Universe this may have been 40\%. Note that this assumes that clusters in the early Universe followed the same mass-radius relation as clusters do today. If they were denser, the 40\% quoted here is a lower limit. Locally, only 10\% of all stars are born in bound clusters \citep{lada03}, whereas at the high densities of the young Universe this may have been up to 50\% \citep{kruijssen12d}.

Combining all the above numbers, we find that in the present-day Universe, only 1\% of all stars are born in an environment of too high density to allow habitable planet formation, whereas at high redshift ($z=2$--$3$) up to 20\% of all stars did. A rough interpolation suggests that in the formation environment of the Solar system ($z\sim0.5$) at least 90\% of all stars resided in sufficiently low-density environments to potentially host habitable planetary systems. This number is representative for the Universe as a whole -- making the highly simplifying assumption that about half of the stars in the Universe formed in present-day galactic environments, whereas the other half formed in a high-density environment, we find that some \emph{10\% of all stars in the Universe may not have planets in their habitable zone due to environmental effects}. The remaining 90\% of stars formed in a setting that in terms of planet formation was as benign as the field. We conclude that while very interesting in its own right, the environmental inhibition of habitable planet formation is likely not important for the cosmic planet population.

\section{\textbf{Linking YMC formation to globular clusters and other major star formation events in the Universe}}
\label{sec:glob_clusters}

Globular clusters (GCs -- extremely dense clusters of old stars) are found in galaxies of all masses, but the most prominent ($N_{\rm GC}>100$) populations exist in galaxies of stellar masses exceeding a few $10^{10}~\msun$ \citep{peng08}. The Milky Way contains over 150 old globular clusters, which have ages of $5$--$13$~Gyr \citep{forbes10} and masses of roughly $10^4$--$10^6~\msun$ \citep{harris96,brodie06}. The detection of YMCs with masses comparable to or greater than GCs in nearby merging and starburst galaxies \citep{schweizer82,holtzman92,schweizer96,whitmore99} has reinvigorated discussion of their origins. Did the most massive YMCs form in a way similar to GCs? If true, this would provide a unique window to understanding extreme cluster formation at or before the peak of cosmic star formation \citep[$z\sim2$--$3$,][]{madau96,hopkins06b,bouwens11,moster13}.

The mass spectrum of GCs is fundamentally different than that of young clusters. While the latter follows a power law ${\rm d}N/{\rm d}M\propto M^{-2}$ over most of its mass range $M=10^2$--$10^8~\msun$ (\citealt{zhang99, larsen09}; except for possibly an exponential truncation at the high-mass end, see e.g.,~\citealt{gieles06b}), GCs have a characteristic mass scale of $M\sim10^5~\msun$ \citep{harris96,brodie06}. The modern interpretation is that GCs initially followed the same mass spectrum as young clusters in the local Universe \citep[e.g.][]{elmegreen97,kravtsov05,kruijssen12b}, but that it subsequently evolved into its present, peaked form. Clusters much more massive than the characteristic GC mass likely existed, but were rare due to the steep slope of the mass spectrum. In addition, the maximum mass of GCs has been influenced by their formation environment (see below) and/or dynamical friction \citep[e.g.][]{tremaine75}. On the other side of the mass spectrum, the numerous low-mass GCs were disrupted and did not survive until the present day \citep{fall01}. However, this explanation is not undisputed. Firstly, the characteristic mass scale of GCs is remarkably universal \citep{jordan07}, whereas the disruption of stellar clusters is sensitive to the galactic environment \citep[e.g.][]{kruijssen11}.  Secondly, in the Fornax dwarf spheroidal galaxy the metal-poor GCs constitute 25\% of the field stellar mass at similar metallicities \citep{larsen12}, which is two orders of magnitude higher than the `universal' baryonic mass fraction of GCs in giant elliptical galaxies found by \citet{mclaughlin99}. This suggests that very specific conditions are required for a full cluster mass spectrum to have been present initially -- in particular, all coeval stars need to have formed in bound clusters \citep[at odds with theory and observations, see e.g.][]{kruijssen12d}, of which the surviving GCs did not lose any mass at all, whereas the lower mass clusters were all completely destroyed. Several adjustments to the simple disruption model have therefore been proposed (see below). For the comparison to YMCs, the key point is that the conditions of GC formation must have been such that they could survive a Hubble time of dynamical evolution, irrespective of their particular formation mechanism(s). This itself can be used to constrain their formation environment.

Early GC formation theories attributed the characteristic mass scale of GCs to specific conditions at the time of GC formation, which historically was mainly driven by the fact that no open clusters were known to have masses comparable to GCs. Many of these theories addressed the problem of producing a sufficient mass concentration of dense gas needed to form a GC. Since the Jeans mass following recombination is $M_{\rm J}=10^5$--$10^6~\msun$, \citet{peebles68} suggested that GCs were the first bound structures to form. Similarly, \citet{fall85} suggested that proto-GC clouds formed though thermal instabilities in galactic haloes from metal-poor gas with a Jeans mass above 10$^6~\msun$. \citet{ashman92} put forward the idea that the shocks in gas-rich galaxy mergers lead to the formation of extremely massive GMCs and GCs,  in line with observations of ongoing mergers in the local Universe \citep[e.g.][]{whitmore99}. Alternatively, it has been proposed that GCs formed in low-mass dark matter haloes, which were subsequently lost due to a Hubble time of tidal stripping and/or two-body relaxation \citep[e.g.][]{bekki08,griffen10}. Finally then, GCs may represent the former nuclear clusters of cannibalised dwarf galaxies \citep{lee09} -- note that in general, the hierarchical growth of galaxies implies that some fraction of the GC population in a galaxy previously belonged to smaller systems \citep[e.g.][]{muratov10}.

All of the above models require that the distribution of GCs should follow the spatial structure of the galactic stellar halo -- either (1) because they were initially directly associated with dark matter, (2) because they formed at very early times, before the main bodies of present-day galaxies were in place, or (3) because their production/accretion took place during hierarchical galaxy growth, which predominantly populates galaxy haloes \citep[e.g.][]{sales07}. It has been shown that metal-rich ($[{\rm Fe}/{\rm H}]>-1$) GCs are associated with the stellar mass in galaxies (in the Milky Way, the metal-rich GC population even exhibits a net rotation of $60~{\rm km}~{\rm s}^{-1}$, see \citealt{dinescu99}), whereas the metal-poor ($[{\rm Fe}/{\rm H}]<-1$) part of the GC population extends further and is associated with the stellar halo \citep{brodie06,strader11b}. The above GC formation scenarios may therefore possibly apply to metal-poor GCs, but are unlikely to hold for the metal-rich ones. Overall though, a re-examination of these models has mainly been prompted by the discovery of the Hubble Space Telescope that YMCs with masses exceeding those of GCs are still forming in the Universe today \citep[e.g.][]{whitmore99}.

The elevated gas accretion rates in high-redshift galaxies lead to extremely turbulent (with Mach numbers ${\cal M}\sim100$), clumpy galaxies, which have high Toomre (or Jeans) masses, allowing the clumps to reach masses up to $10^9~\msun$\citep{elmegreen05,forsterschreiber09,krumholz10,forbes12}. In these conditions, it seems plausible that the high stellar masses required for GC progenitors can easily be assembled \citep{shapiro10}, and considering the high gas densities this is likely accompanied by locally highly efficient star formation and hence a negligible effect of any disruption due to gas expulsion (see \S~\ref{sub:theory}). The formation of gravitationally bound, extremely massive clusters seems inevitable in the high-redshift Universe -- examples of similar conditions in the nearby Universe exist too \citep{schweizer82,holtzman92,miller97,bastian06}. While these examples mainly refer to galaxy mergers, which are not thought to be the main source of star formation in the early Universe \citep[e.g.][]{genzel10}, the conditions in the interstellar medium are similar \citep{kruijssen13d}. Importantly though, these nearby examples are all accompanied by a population of low-mass clusters \citep[see above and e.g.][]{zhang99}.

Despite the fact that GCs presently reside in widely differing galactic environments, their common ancestry may provide an explanation for their characteristic mass-scale, even if they formed in the same way as clusters in nearby galaxies. The high-mass end of the cluster mass function is likely truncated by the Toomre mass, which is the largest mass-scale in a galaxy over which gas clumps can become self-gravitating and ranges from several $10^6~\msun$ in Milky Way-like spirals to several $10^9~\msun$ in extreme, high-redshift environments. For a global star formation efficiency of $10\%$ (appropriate for the $\sim100$-pc scales considered here), this gives maximum cluster masses between several $10^5~\msun$ and several $10^8~\msun$. Evidence for such a truncation exists in observations of nearby galaxies \citep{gieles06b,larsen09} and it has also been derived indirectly from the present-day GC mass spectrum \citep{jordan07,kruijssen09b}. As discussed above, the high initial masses required for GCs greatly constrain their possible formation environment to the vigorously star-forming, highly turbulent environments seen at high redshift where the Toomre mass is sufficiently high to form proto-GCs ($M_{\rm T}\sim10^8~\msun$, see below). These galaxies are characterized by high accretion rates and densities, and are forming large numbers of clusters per unit time, leading to the formation of extremely massive clusters (see \citealt{bastian08} for local examples). However, these high ambient densities also lead to very efficient cluster disruption -- predominantly due to tidal shocking \citep{spitzer58,gieles06}. This suggests a picture in which the vast majority of GC disruption occurred at high redshift, removing the low-mass clusters from the GC population and leaving only the most massive ones intact \citep{elmegreen10,kruijssen12c}.

Given a high-mass truncation of the initial cluster mass function, the destruction of low-mass clusters eventually leads to a saturation of the characteristic mass-scale at $\sim10\%$ of the truncation mass \citep{gieles09}. For a Toomre mass of $M_{\rm T}\sim10^8~\msun$, a global star formation efficiency of $10\%$ (as assumed above) gives an initial truncation mass of $M_{\rm max,i}\sim10^7~\msun$. A key requirement is that the resulting GCs survived until the present day and hence at some point decoupled from their high-density natal environment. In the most extreme environments, cluster disruption may typically remove up to $90\%$ of the mass from massive clusters before the disruption rate starts decreasing due to cluster migration \citep{kruijssen11}, and hence the truncation mass would have decreased to a few $10^6~\msun$ by the time the most massive GCs escaped to a more quiescent environment (such as the galaxy halo). At $10\%$ of the truncation mass, this implies a {\it universal} characteristic mass-scale of GCs of a few $10^5~\msun$, which is indeed what is seen in the Universe today. In this interpretation, the mass-scale of GCs indirectly reflects the Toomre mass at the time of their formation -- much like the masses of YMCs do in nearby galaxies.
 
A fundamental difficulty in trying to reconstruct the conditions of GC formation from the present-day GC populations that exist in nearby galaxies, is that the evolution of dense stellar systems with two-body relaxation times shorter than a Hubble time is convergent \citep{gieles11b}. In other words, the characteristics imprinted by their formation process will be washed out by stellar dynamics. It is therefore necessary to address the formation of GCs from more direct avenues.

During the last decade, the discovery of multiple main sequences and chemical abundance variations in GCs (see \citealt{gratton12} for a review) has triggered a re-evaluation of GC formation mechanisms, suggesting that they formed during multiple star formation episodes (spaced by 10--100~Myr) and therefore underwent chemical self-enrichment \citep[e.g.][]{conroy11}. These new results have even led to the idea that GCs formed by a fundamentally different process than YMCs in the nearby Universe. However, without any direct observations to support such a far-reaching interpretation, it remains conceivable that `normal' YMC formation at the low metallicities and high densities that characterize the high-redshift Universe naturally lead to the observed properties of GCs.

There exist no nearby YMC-forming regions of adequately low metallicity ($[{\rm Fe}/{\rm H}]=-1.7$ to $-0.7$) {\it and} high mass ($M_{\rm YMC}>10^6~\msun$) to probe the formation of GC-like YMCs. Hence, the question is to what extent nearby YMCs are representative of young GCs. No indication of ongoing star formation has been detected in massive ($M=10^4$--$10^8~\msun$) and young-to-intermediate-age ($\tau=10$--$1000$~Myr) clusters in nearby galaxies \citep{bastian13}, suggesting that these all formed in a single burst of star formation. A recently proposed model by \citet{bastian13b} argues that the high stellar densities of young GCs imply that stellar winds and binary ejecta would have been efficiently swept up by the long-lived protostellar disks around low-mass stars. Because low-mass stars are still fully convective in the pre-main sequence phase, they are then uniformly polluted. In this `early disk accretion' model, only the stars that passed through the core of the cluster during the right time interval exhibit abundance variations, naturally leading to relatively bimodal abundance variations and the (incorrect) impression of multiple episodes of star formation. The model will need to be verified in future work, but for now it illustrates that GC formation may very well have proceeded in a way that is consistent with the known physics of YMC formation.

Given the available evidence, it seems plausible that at least the metal-rich part of the GC population formed in a way similar to YMCs in nearby galaxies. The study of YMC formation can therefore provide key insights in the formation of the oldest structures in the local Universe. However, the extended spatial distribution of metal-poor clusters with respect to the galaxy light \citep{brodie06} suggests that a different formation mechanism may have been at play. Tentative support is provided by the lack of field stars with metallicities similar to the (metal-poor) GCs in the dwarf spheroidal galaxy Fornax \citep{larsen12} -- if these GCs formed like YMCs as part of a full mass spectrum of lower-mass clusters that were subsequently disrupted, a larger population of coeval field stars would be expected. Fornax is an extreme case though -- for more massive galaxies the metal-poor and metal-rich GC sub-populations may emerge naturally in the context of hierarchical galaxy formation \citep{muratov10,tonini13}. Studies of the co-formation of galaxies and their cluster populations will constrain GC formation further in the coming years.

\section{Summary and outlook}

We conclude that the study of YMC formation will play a key role in our future understanding of star and planet formation across cosmological timescales. YMCs act as a natural bridge linking what we learn from star-forming regions in the solar neighborhood, to starburst systems in the Local Universe, globular cluster formation and star/planet formation at the earliest epochs of the universe.

In the same way that the discovery and characterization of YMCs with Hubble and 8-m-class optical/infrared telescopes revolutionized our understanding of the relationship between open and globular clusters, the advent of new survey data and new/upcoming facilities (ALMA, VLA, JWST, GAIA, ELTs) are set to revolutionize our understanding of the YMC formation process. In five to ten years we should hopefully have a complete sample of YMC progenitor clouds in the Galaxy and demographics of YMC progenitors and YMCs in a large number of external galaxies, with a wide range of environments (mergers/starbursts, low-metallicity systems, centers of galaxies etc.). With this data it will be possible to determine (i) the global environmental conditions required to assemble gas to such high density, (ii) if YMCs in different environments assemble their mass differently, (iii) how this is related to the environmental variation of the critical density for star formation, and (iv) if the properties of stars and planets in YMCs are affected by differences in the initial protostellar density.

The fact that for the foreseeable future the Milky Way is the only place it will be possible to resolve individual (proto)stars means Galactic studies will be crucial. Follow-up studies connecting the initial gas conditions to the subsequent stellar populations will be able to probe the assembly of stellar and planetary mass as a function of environment. GAIA and complementary, follow-up spectroscopic surveys will nail down the 6D stellar structure of the nearest clusters, unambiguously measuring any primordial rotation or expansion as well as any dynamical evidence for sub-clusters. This will directly answer several of the key questions outlined in this review.

We conclude that the next few years are set to be an exciting and productive time for YMC studies and will likely lead to major breakthroughs in our understanding of YMC formation.

\bigskip

\textbf{ Acknowledgments.} We would like to thank Giacomo Beccari, Henrik Beuther, Yanett Contreras, Andreea Font, Roberto Galvan-Madrid, Jens Kauffmann, Mark Krumholz, Betsy Mills, Fr\'ed\'erique Motte, Cornelia Lang, Quang Nguyen Luong, Thushara Pillai, James Urquhart, Qizhou Zhang and the anonymous referee for helpful comments on the manuscript. We acknowledge the hospitality of the Aspen Center for Physics, which is supported by the National Science Foundation Grant No. PHY-1066293.

\bibliographystyle{ppvi_lim1}
\bibliography{mybib}

\end{document}